\documentclass[
    a4paper,
    prl,
    twocolumn,
    superscriptaddress,
    floatfix
]{revtex4-1}

\usepackage{amsmath,amssymb}
\usepackage{lipsum,soul}
\usepackage{bm}
\usepackage{graphicx}
\usepackage{graphics}
\usepackage{amsmath}
\usepackage{array}
\usepackage[caption=false]{subfig}
\usepackage{xcolor}
\usepackage{subfig}
\usepackage{hyperref}
\usepackage[capitalise]{cleveref}
\usepackage{textcomp} 

\usepackage{epstopdf}
\usepackage{cases}
\usepackage{amssymb}
\usepackage{multirow}
\usepackage{siunitx}
\usepackage{cases}
\usepackage{tabularx}
\usepackage{nomencl}
\usepackage[utf8]{inputenc}
\usepackage{tikz}
\usepackage[T1]{fontenc}
\usepackage{comment}
\usepackage{lineno,hyperref}
\modulolinenumbers[5]
\usepackage{booktabs}

\begin{document}

\title{
Violation of Fourier's law in homogeneous systems}
\author{Chuang Zhang}
\affiliation{State Key Laboratory of Coal Combustion, School of Energy and Power Engineering, Huazhong University of Science and Technology,Wuhan, 430074, China}
\author{Dengke Ma}
\affiliation{NNU-SULI Thermal Energy Research Center (NSTER) and Center for Quantum Transport and Thermal Energy Science (CQTES), School of Physics and Technology, Nanjing Normal University, Nanjing, 210023, China}
\author{Manyu Shang}
\affiliation{School of Physics and Wuhan National High Magnetic Field Center, Huazhong University of Science and Technology, Wuhan 430074, P. R. China}
\author{Xiao Wan}
\affiliation{State Key Laboratory of Coal Combustion, School of Energy and Power Engineering, Huazhong University of Science and Technology,Wuhan, 430074, China}
\author{Jing-Tao L\"u}
\affiliation{School of Physics and Wuhan National High Magnetic Field Center, Huazhong University of Science and Technology, Wuhan 430074, P. R. China}
\author{Zhaoli Guo}
\email{Corresponding author: zlguo@hust.edu.cn}
\affiliation{State Key Laboratory of Coal Combustion, School of Energy and Power Engineering, Huazhong University of Science and Technology,Wuhan, 430074, China}
\author{Baowen Li}
\email{Corresponding author: Baowen.Li@Colorado.Edu}
\affiliation{Paul M Rady Department of Mechanical Engineering, Department of Physics, University of Colorado, Boulder, Colorado 80309, USA}
\author{Nuo Yang}
\email{Corresponding author: nuo@hust.edu.cn}
\affiliation{State Key Laboratory of Coal Combustion, School of Energy and Power Engineering, Huazhong University of Science and Technology,Wuhan, 430074, China}

\date{\today}

\begin{abstract}

Hotspot is a ubiquitous phenomenon in micro/nanoscale chips. Here, it is found that Fourier’s law is invalid in such a homogeneous system. The hotspots in homogeneous 2D disk/3D sphere and graphene disk are studied based on phonon Boltzmann transport equation. Instead a constant value, a graded thermal conductivity is observed. The mechanisms of phonon scattering are analyzed. It is found that for a system with fixed size, the graded thermal conductivity is predictable as long as there is not sufficient phonon scattering, which is independent on material properties, dimensions or system size. This work may shed light on both theoretical and experimental studies on heat dissipation.

\end{abstract}
\maketitle

$Introduction.$—Thermal conductivity, a fundamental physical property of materials,
{\color{black}{is a constant that independent of system size and geometry in bulk materials. It is an intrinsic property that depends only on the component of materials.  Heat conduction in such materials generally follows the Fourier's law, which implies that the heat carriers (phonons) undergo a diffusive process~\cite{kaviany_2008,ChenG05Oxford}.
However, as the size or dimension of the system decreases, in particular when the size goes down to nanoscale and/or dimension is reduced to two dimension (2D) or quasi-one dimension (1D), there is still no rigorous mathematical proof that the Fourier's law is still valid. In contract, many researchers discovered that the thermal conductivity is a function of size and geometry~\cite{ZHANG2020,ballisticsiliconapl2001,RevModPhysLibaowen,RevModPhys.90.041002,li_thermal_2003,hsiao_observation_2013,PhysRevLett.118.135901,chang_breakdown_2008,xu_length-dependent_2014,PhysRevB.38.1963,chen1996,PhysRevB.90.064302}.}}

The {\color{black}{underlying physical mechanisms of non-Fourier heat conduction}} mainly include: {\color{black} First, when the size of structures is comparable with the phonon mean free path, the phonon transport is largely affected by the boundary scattering~\cite{MajumdarA93Film,hsiao_observation_2013},
such that the thermal conductivity can be altered significantly by nanoengineering~\cite{volz1999,li_thermal_2003,hsiao_observation_2013,PhysRevLett.118.135901,chang_breakdown_2008,xu_length-dependent_2014}.}
Second, a divergent thermal conductivity with system size was found in many low dimensional momentum conserved systems because of the existence of zero frequency and large wave length modes
~\cite{PhysRevLett.84.2857,narayan_anomalous_2002,lepri_thermal_2003,Dhar2008advphys}. {\color{black} Third, as the system size is close or comparable to the phonon wave length, the wave nature of phonons is non-negligible in thermal transport~\cite{PhysRevB.67.195311,ma2019b}.}
{\color{black}{Fourth, the possible existence of the second sound  makes heat transfer like wave propagation~\cite{PhysRev_GK,PhysRev.148.766,leesangyeopch1,lee_hydrodynamic_2015,cepellotti_phonon_2015,WangMr15application,huberman_observation_2019}. This regime is usually called phonon hydrodynamic regime.}}

Most studies so far have focused on the length-dependent thermal conductivity~\cite{ZHANG2020,RevModPhys.90.041002,lepri_thermal_2003,Dhar2008advphys,hsiao_observation_2013,PhysRevLett.118.135901,chang_breakdown_2008,xu_length-dependent_2014}.
The quasi-ballistic thermal transport effects~\cite{PhysRevB.97.014307,PhysRevApplied.10.054068} are also measured with a nanoscale heat source comparable to the
phonon mean free path~\cite{ballisticsiliconapl2001,PhysRevLett.107.095901,hu_spectral_2015,siemens_quasi-ballistic_2010}. The difference from Fourier’s law is just the value of thermal conductivity depends on the size of heat source.
Thermal conductivity, defined through the Fourier's law, is homogeneous in nanostructures~\cite{hu_spectral_2015,siemens_quasi-ballistic_2010}.

Recently, in a system with fixed size, an abnormal phenomenon - graded thermal conductivity - the thermal conductivity in the
radial direction increases with the distance from the disk center, has been observed in homogeneous nanoscale graphene disk and carbon nanocone by molecular dynamics simulations~\cite{yang_nanoscale_2015,ma_unexpected_2017}.

Due to the limitation of computational resources, the diameter of system in previous molecular dynamics simulations~\cite{yang_nanoscale_2015,ma_unexpected_2017} is below $25$ nanometers.
Does the graded thermal conductivity exist in a macro-system?
What is the physical understanding on the mechanisms of
graded thermal conductivity in homogeneous system?

\begin{figure}[htb]
     \centering
    \includegraphics[scale=0.18,clip=true]{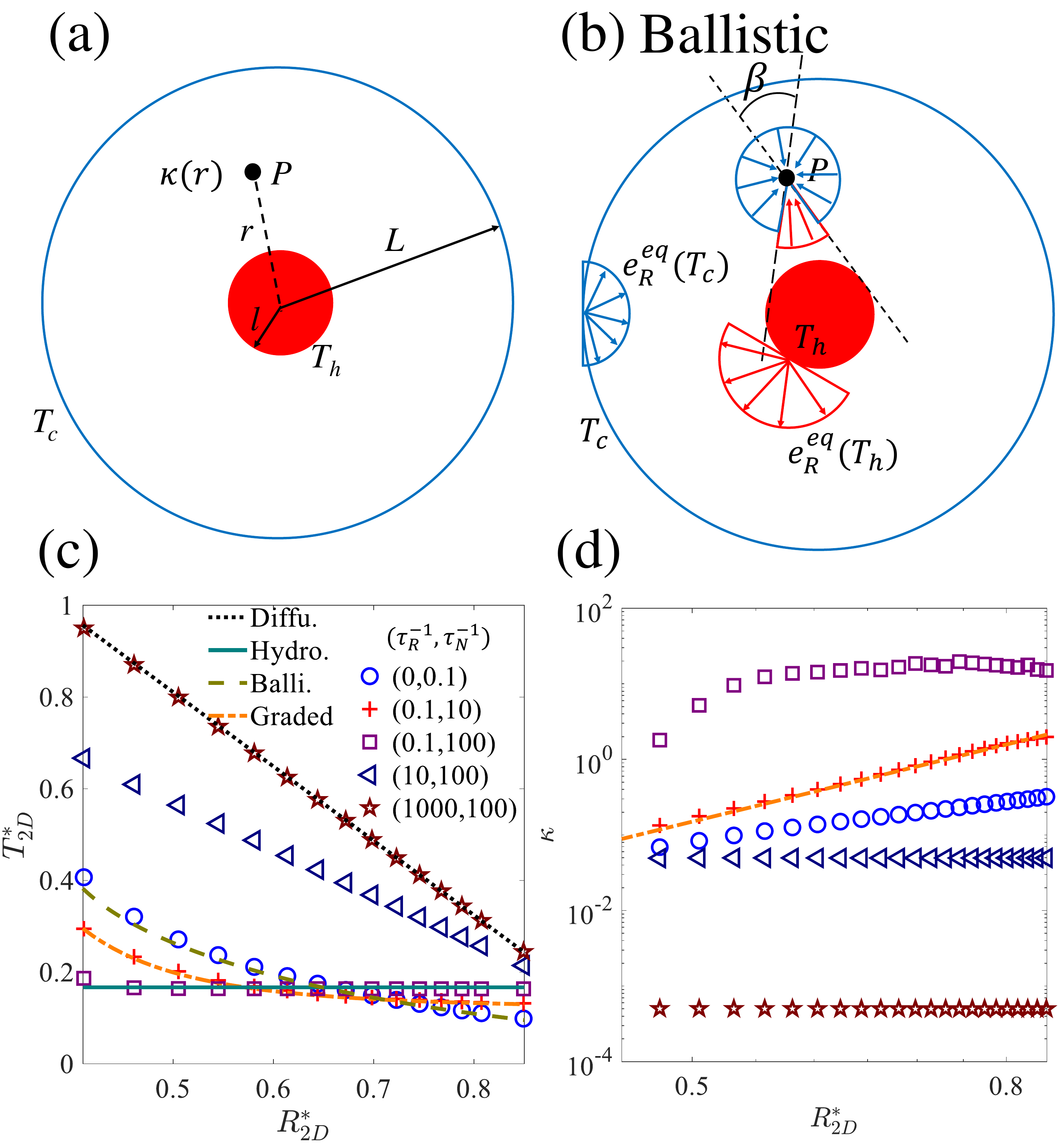}
     \caption{
     (a) Schematic of homogeneous 2D disk. (b) Asymmetric phonon transport in ballistic limit. For a fixed system size $L=5l=0.5$, the temperature profile (c) and graded thermal conductivity (d) along the radial direction with different scattering rates, where $T_{2D}^*= (T-T_c)/ \Delta T$, $R_{2D}^* = (\ln(r/l) -1 ) /( \ln(L/l) -1)$, $r$ is the distance from the disk center. More results and details are shown in SM IV and TABLE. S1.
     }
     \label{2DD}
\end{figure}

\begin{figure}[htb]
     \centering
   \includegraphics[scale=0.19,clip=true]{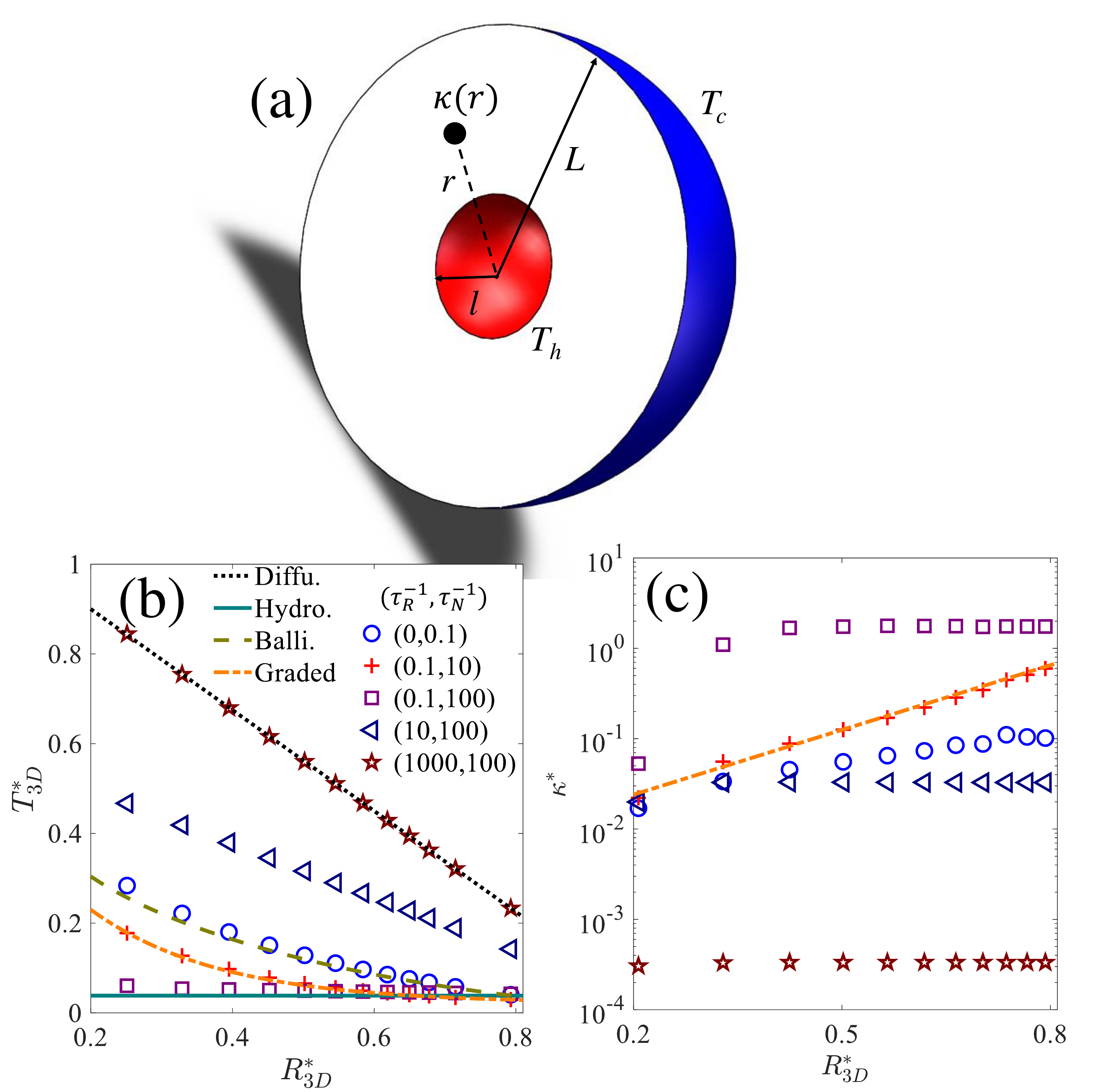}
     \caption{(a) Schematic of 3D sphere. For a fixed system size ($L=5l=0.5$), the temperature profile (b) and graded thermal conductivity (c) along the radial direction with different scattering rates, where $T_{3D}^*= (T-T_c)/ \Delta T$, $R_{3D}^* = (1/l -1/r +1 ) / (1/l -1/L +1 ) $, $r$ is the distance from the ball center. More results and details are shown in SM IV and TABLE. S2.
     }
     \label{3DD}
\end{figure}

\begin{figure}[htb]
 \centering
 \includegraphics[scale=0.32,clip=true]{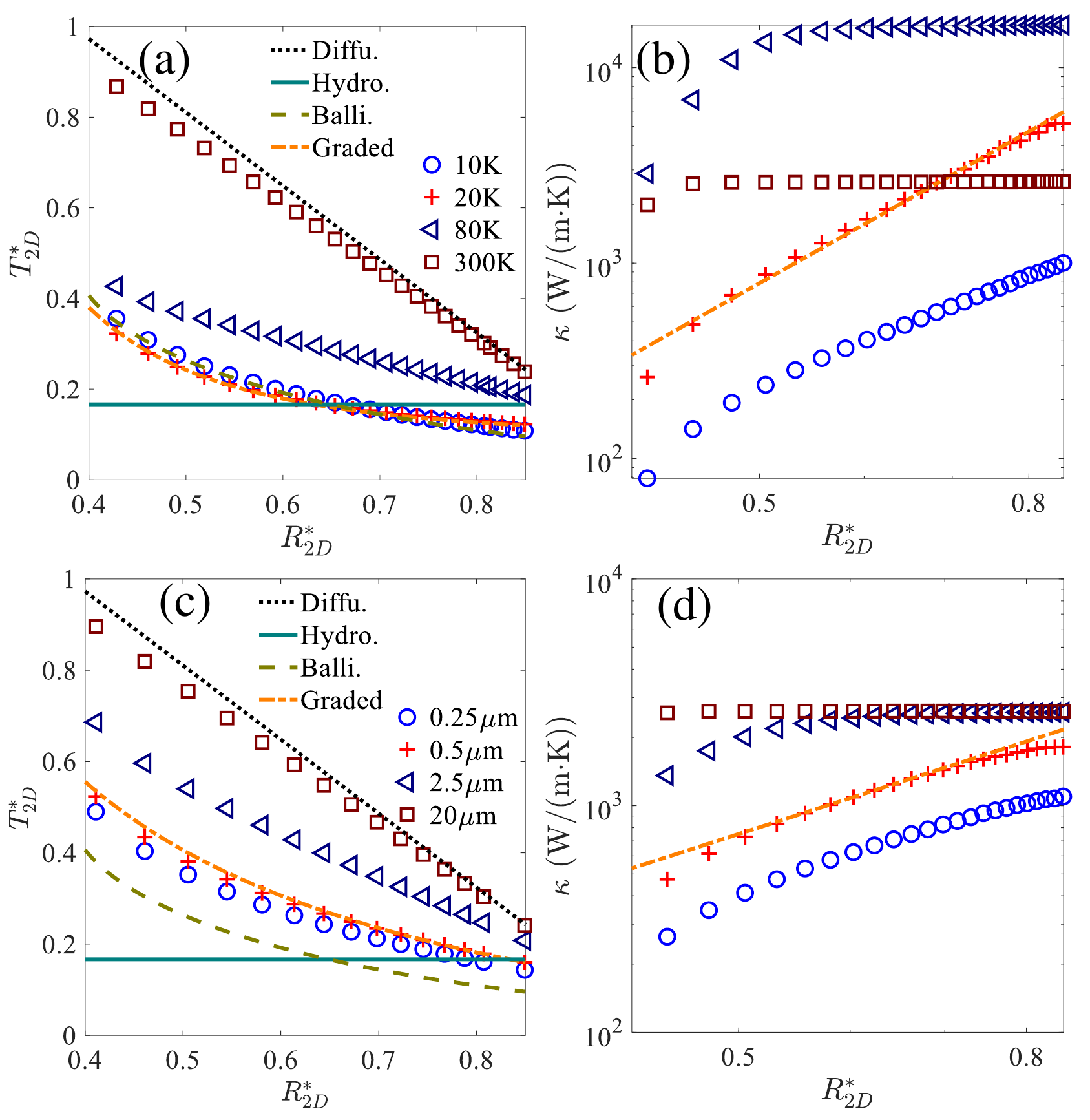}
 \caption{The temperature profile and thermal conductivity in graphene disk~\cite{yang_nanoscale_2015}. (a)(b) Fixed size $L=40~\mu$m. (c)(d) Fixed $T_0=300~\text{K}$. More results and details are shown in SM IV and TABLE. S3 and S4.
 }
 \label{diskGD}
\end{figure}

{\color{black}{In this Letter, we shall answer above mentioned questions by studying the graded thermal conductivity in homogeneous 2D disk/3D sphere (\cref{2DD,3DD}) with a fixed macroscopic size from ultra-low temperature to high temperature, without limiting to any specific material.
The underlying physical mechanisms are to be analyzed by ballistic phonon transport, normal (N) scattering and resistive (R) scattering, respectively.
The general conclusion will be exemplified by the graphene disk (\cref{diskGD}).}}

The schematics of the 2D disk and 3D sphere are shown in~\cref{2DD}(a) and~\cref{3DD}(a), respectively, where the radii of the inner and outer heat baths are $l$ and $L$, respectively.
The temperatures of the inner and outer heat baths are fixed at $T_h=T_0+\Delta T/2$ and $T_c=T_0 -\Delta T/2$, where $\Delta T/T_0 \rightarrow 0^{+}$.
The local radial thermal conductivity $\kappa$ is calculated by
\begin{equation}
\kappa (r)= -\frac{q(r)} {dT/dr } , \quad l <  r  < L,  \label{eq:Qkr}
\end{equation}
where $q(r)$ is the local heat flux, namely the heat energy flow along the radial direction per unit area in a unit time. $T(r)$ is the local temperature, $r$ is the distance from the center, 

$Model~equation.$—We start with the steady-state phonon Boltzmann transport equation (BTE) under the Callaway model and Matthiessen's rule~\cite{ChenG05Oxford,PhysRev_callaway,wangmr17callaway,PhysRevB.97.134307,luo2019}, in which both the normal (N) scattering and resistive (R) scattering are included.
\begin{align}
\bm{v}  \cdot \nabla_{\bm{x}} e = \tau_R^{-1} (e^{eq}_{R}-e ) + \tau_N^{-1} (e^{eq }_{N}-e ), \label{eq:BTE}
\end{align}
where $e$ is the phonon distribution function of energy density, $\bm{v}$ is the group velocity, $\bm{x}$ is the spatial position.
The heat flux and temperature in Eq.~\eqref{eq:Qkr} are obtained by taking the moment of the distribution function.
$e^{eq}_{R}$ and $e^{eq}_{N}$ are the associated phonon equilibrium distribution functions of energy density for R and N scattering, respectively.
$\tau_R$ and $\tau_N$ are the relaxation times for R and N scattering, respectively.
In the BTE simulations~\cite{li2019,ChenG05Oxford,wangmr17callaway,luo2019}, the wave nature of phonons is not taken into account~\cite{PhysRevB.67.195311,ma2019b}.
{\color{black}{The distribution functions of all phonons emitting from the inner (or outer) heat bath are $e_R^{eq} (T_h)$ (or $e_R^{eq} (T_c)$)~\cite{wangmr17callaway,luo2019}.}}
More details of phonon BTE and boundary conditions is shown in Supplemental Material(SM) I.

The phonon transport will be simulated by solving phonon BTE numerically by the implicit discrete ordinate method~\cite{wangmr17callaway,ZHANG20191366}.
In simulation of 2D disk/3D sphere with a fixed macroscopic size, the Debye approximation and gray model~\cite{ChenG05Oxford} are used, where no phonon dispersion and polarization are considered.
{\color{black}{Note that the heat conduction in 2D disk/3D sphere is not limited by specific materials properties so that all physical variables are dimensionless.
The radii of inner and outer heat baths are fixed at $l=0.1$ and $L=0.5$, respectively (\cref{2DD,3DD}).
The group velocity is $|\bm{v}|=1$ and the specific heat is $C=1$.}}
The thermal effects of N (R) scattering on graded thermal conductivity will be investigated by adjusting the values of $\tau_{N}^{-1}$ or $\tau_{R}^{-1}$.

In simulation of graphene disk (\cref{diskGD}), {\color{black}{the phonon dispersion and polarization of graphene are calculated using Vienna Ab initio Simulation Package (VASP) combined with phonopy.}}
And the effects of both frequency-dependent N and R scattering will be considered.
More details on phonon properties of graphene and numerical solutions can be found in SM II-III.

$Results.$—The phonon transport in a homogeneous 2D disk {\color{black}{with a fixed macroscopic system size}} is studied first.
In addition to numerical results, the analytical solutions in the ballistic~\cite{olfe1968,li2019} ($\tau_{R}^{-1}= 0,~\tau_{N}^{-1}=0$), diffusive ($\tau_{R}^{-1} \rightarrow \infty ,~\tau_{N}^{-1}=0$) and hydrodynamic~\cite{PhysRev_GK,PhysRev.148.766,Nanalytical,lee_hydrodynamic_2015,cepellotti_phonon_2015}  ($\tau_{N}^{-1} \rightarrow \infty,~\tau_{R}^{-1}=0$) limits are also plotted in~\cref{2DD} to show the separate thermal effects of N or R scattering (Derivations of three limits are shown in SM IV).

At ultra-low temperature,  phonon-phonon interaction/scattering can be totally neglected and ballistic phonon transport dominates heat conduction~\cite{MajumdarA93Film,li2019} (e.g., $\tau_R^{-1}=0,~\tau_N^{-1}=0.1$).
As shown in~\cref{2DD}(c)(d), the temperature profile is nonlinear and the radial thermal conductivity is not a constant anymore, instead it depends on $r$.
The results are consistent with the analytical solutions in the ballistic limit~\cite{olfe1968,li2019} (see~\cref{2DD}(b) or SM IV), i.e.,
\begin{align}
T(r) &= \frac{2\arcsin (l/r) }{2\pi} T_h + \left(1-\frac{2\arcsin (l/r) }{2\pi}\right) T_c. \label{eq:2dTballistic}
\end{align}
This suggests the graded thermal conductivity, similar to what was observed in nanodisks~\cite{yang_nanoscale_2015} and nanocones~\cite{ma_unexpected_2017} by molecular dynamics simulations.

At low temperature, R scattering is weak and N scattering dominates the heat conduction so that phonon transports in the phonon hydrodynamic regime~\cite{PhysRevB.99.085202,PhysRev_GK,PhysRev.148.766,Nanalytical,WangMr15application,lee_hydrodynamic_2015,cepellotti_phonon_2015}.
It can be observed that with the increase of $\tau_N^{-1}$, the slopes of the numerical profiles of graded thermal conductivity in~\cref{2DD}(d) increase first and then decrease gradually.
As N scattering is much stronger than R scattering, the radial temperature goes to a constant and recovers the phonon hydrodynamic limit~\cite{PhysRev_GK,PhysRev.148.766,Nanalytical,lee_hydrodynamic_2015,cepellotti_phonon_2015} (see SM IV), i.e.,
\begin{align}
T =\frac{lT_h+L T_c } {l+L}. \label{eq:2dThydrodynamic}
\end{align}

At high temperature, R scattering starts to dominate the heat conduction so that phonon transport goes to the diffusive regime.
It can be observed that with the increase of $\tau_R^{-1}$, the temperature profile comes to linear and the graded thermal conductivity phenomenon disappears.
The results agree well with the analytical solutions in the diffusive limit (see SM IV), i.e.,
\begin{align}
dT &\propto  d \ln r. \label{eq:2dTdiffusive}
\end{align}

$Physical~mechanisms.$—As shown in~\cref{2DD}, in homogeneous 2D disk with a fixed macroscopic size, the non-Fourier's thermal transport phenomenon depends on scattering, i.e., $\tau_N^{-1}$ and $\tau_R^{-1}$.
In the following, the underlying physical mechanisms of phonon scattering are discussed in details.

$Ballistic.$—In the ballistic regime, corresponding to ultra-low temperature, phonon-phonon interaction/scattering rarely exists.
Phonon advection dominates heat conduction~\cite{PhysRevB.99.085202,li2019}.
For any point in the interior domain, phonons reach this point from the inner and outer thermal baths with different directions~\cite{chen1996,li2019}.
Both analytical (Eq.~\eqref{eq:2dTballistic}) and numerical results predict that the temperature profile along radial direction has a non-linear dependence on the distance $r$ in 2D disk (\cref{2DD}(c)).
It is different from ballistic phonon transport in a symmetric system, in which the temperature is a constant~\cite{MajumdarA93Film}.
In the symmetric system, all phonons emitting from one heat bath will be totally received by the other (see FIG. S2).
So that the temperature gradient inside the system vanishes~\cite{MajumdarA93Film}.

For ballistic transports in 2D disk, all phonons emitting from the inner bath will be received by the outer bath.
However, phonons emitting from the outer heat bath will be received by both the inner and outer heat baths (see~\cref{2DD}(b)).
That means a portion of phonons are not received by the inner bath, which do not contribute to heat flux, but contribute to local energy or temperature.
The temperature gradient is built by the asymmetric phonon advection, instead of phonon-phonon scattering.
Because the heat flux from inner to outer is conserved, graded thermal conductivity can be observed in 2D disk in the ballistic regime (\cref{2DD}(d)).

$Scattering.$—With the increase of temperature, phonon-phonon scattering becomes strong and dominates heat transfer~\cite{PhysRevB.99.085202}.
In this case, the thermal effects of N scattering ($\tau_N^{-1}$, momentum conserved) and R scattering ($\tau_R^{-1}$, momentum not conserved) are discussed as follows.

$N ~scattering.$—At low temperature, R scattering is weak and N scattering dominates the phonon transport.
N scattering does not cause thermal resistance~\cite{PhysRev_GK,PhysRev.148.766,Nanalytical,WangMr15application,lee_hydrodynamic_2015,cepellotti_phonon_2015}, but affects energy distribution and temperature profile.
When N scattering is weak $( \tau_N^{-1} \leq 10)$, the graded thermal conductivity is attributed to asymmetric scattering.
This means that N scattering is frequent far from the center.
But near the center, the N scattering is less, which limits the exchange of thermal energy.
As $\tau_N^{-1} \gg 10$, the N scattering is very strong inside the whole domain and goes to a constant temperature profile~\cite{Nanalytical,PhysRev_GK,PhysRev.148.766} (Eq.~\eqref{eq:2dThydrodynamic}), namely, graded thermal conductivity disappears.
In a word, it can be observed that as N scattering increases, the graded thermal conductivity phenomenon increases first, and then fades away (\cref{2DD}(d)).

$R ~scattering.$—At high temperature, R scattering starts to play the leading role on heat conduction.
Different from N scattering, R scattering does not conserve momentum, and causes thermal resistance~\cite{kaviany_2008,PhysRevB.99.085202}.
With the increase of R scattering, the frequent energy exchange and heat dissipations decrease the temperature jump near the heat baths~\cite{li2019} (\cref{2DD}(c)).
As $\tau_R^{-1} \gg 10$, the heat conduction follows Fourier’s law and there are a linear temperature profile and a constant thermal conductivity (\cref{2DD}(d)).
In other words, in a structure with frequent R scattering, no graded thermal conductivity appears.

$Graded$ $thermal$ $conductivity.$—{\color{black}{Motivated by previous studies~\cite{yang_nanoscale_2015,ma_unexpected_2017}, an experimental formula of graded thermal conductivity is used to fit the numerical data approximately (\cref{2DD}(d)), i.e.,
\begin{align}
\kappa (r) &=\kappa_0 \left( R_{2D}^*  \right)^{\alpha}, \quad  & R_{2D}^* =\frac{ \ln(r/l)  -1 }{ \ln(L/l) -1 },  \label{eq:2dpowerlaw}
\end{align}
where $\kappa_0 $ is a constant, $R_{2D}^*$ is the normalized coordination in 2D disk and $\alpha$ is the graded rate~\cite{yang_nanoscale_2015,ma_unexpected_2017}.
The detailed fitting parameters can be found in SM IV and TABLE. S1.}}
So that for a fixed disk size, there is no more homogenous value of thermal conductivity, instead a graded increasing thermal conductivity from the disk center to the outer.

For a given 2D disk with a fixed macroscopic size, the above results (\cref{2DD}) and analysis show that the graded thermal conductivity depends on the amount of phonon scattering.
When phonon scattering is not sufficient and $\tau_N^{-1}$ and $\tau_R^{-1}$ are small, the temperature profiles are nonlinear and graded thermal conductivity appears.
{\color{black}{In the ballistic regime, the graded thermal conductivity is caused by the asymmetric phonon advection~\cite{olfe1968,li2019} due to the spatial asymmetry of 2D disk as mentioned in~\cref{2DD}(b) (or FIG. S2) and preceding paragraph (see $ballistic$).
As the phonon-phonon scattering increases, the energy and momentum exchange among phonons break the asymmetric phonon advection gradually.
However, the effects of N and R scattering on graded thermal conductivity are quite different (\cref{2DD}(d)).
With the increase of $\tau_N^{-1}$ (see $N~Scattering$), the graded phenomenon is enhanced first and then fades away due to diverging thermal conductivity~\cite{Nanalytical,PhysRev_GK,PhysRev.148.766}.
As $\tau_R^{-1}$ increases (see $R~Scattering$), the graded phenomenon fades away gradually.}}

$3D~ball.$—Does the graded thermal conductivity exist in 3D structures with a fixed macroscopic size?
In order to look for the answer, the radial thermal conduction in a 3D sphere (\cref{3DD}(a)) is also investigated.
The numerical results in different regimes are shown in~\cref{3DD}(b)(c).
It is found that the temperature profiles and thermal conductivities in 3D sphere are similar to those in the 2D disk.
In addition, an exponential function of graded thermal conductivity is also used to fit the numerical data approximately (\cref{3DD}(c)), i.e.,
\begin{align}
\kappa (r) &=\kappa_0 \exp \left(  \gamma R_{3D}^*  \right) ,  \quad  & R_{3D}^* = \frac{ 1/l -1/r +1  }{1/l -1/L +1 } , \label{eq:3dpowerlaw}
\end{align}
where $R_{3D}^*$ is the normalized coordination in 3D sphere and $\gamma$ is a coefficient.
The detailed fitting parameters can be found in SM IV and TABLE. S2.
Therefore, the graded thermal conductivity can appear in both 2D and 3D radial homogeneous systems with fixed sizes.


{{Besides, the dimensional analysis~\cite{barenblatt1987dimensional} and more results of 2D disk/3D sphere are shown in SM VI.}}

$Graphene~disk.$—Graphene, a very excellent thermal conductor that has been studied extensively~\cite{balandin_superior_2008,RevModPhys.90.041002,yang_nanoscale_2015}, is used to illustrate our analysis.

Firstly, the size of graphene disk is fixed at $L=20~\mu$ m and $l=4~\mu$ m.
Then, the temperature $T_0$ is decreased gradually, as shown in~\cref{diskGD}(a)(b).
As the temperature is changed from $300~\text{K}$ to $3~\text{K}$, it can be observed that graded thermal conductivities and non-Fourier's phenomenon happen, which can be explained that the R scattering becomes weak, and the N scattering starts to dominate the heat transfer~\cite{cepellotti_phonon_2015,lee_hydrodynamic_2015,wangmr17callaway} as the temperature decreases.
At $T_0=30~\text{K}$, the normalized temperature near the inner heat bath is even smaller than that in the ballistic limit, which is impossible if N scattering is weak.
At ultra-low temperature $(3~\text{K})$, ballistic phonon transport dominates heat conduction so that the temperature profile recover the analytical solutions in the ballistic limit.

Secondly, the temperature of graphene disk is fixed at $300~\text{K}$.
Then, the system size $L$ is decreased, as shown in~\cref{diskGD}(c)(d), where $L=5l$.
It can be observed that as system size decreases, the thermal conductivity along the radial direction is not a constant.
Because as system size decreases, the ballistic phonon transport starts to play an important role on heat conduction~\cite{ballisticsiliconapl2001,PhysRevB.99.085202,bae2013ballistic}. It is noted that as the size of graphene disk is tens of nanometers, the graded thermal conductivity has been predicted by molecular dynamics~\cite{yang_nanoscale_2015,ma_unexpected_2017}, which is beyond the applications of phonon BTE~\cite{ChenG05Oxford}.
According to the results of graphene disk (\cref{diskGD}), it can be concluded that graded thermal conductivity occurs at low temperature or for a small sized system, which are consistent with the results in 2D disk/3D sphere.

$Summary ~ and~ Conclusion.$—The thermal conductivity in homogeneous 2D disk/3D sphere and graphene disk with a spot heat source at the center is studied from the phonon Boltzmann transport equation. The results show that, for a homogenous system with fixed size, as long as phonon scattering is not sufficient, the thermal conductivity becomes inhomogeneous, namely, it increases from the center to the outer. 
This study may inspire a better understanding thermal transport in structures with hotspots.

Supported by National Natural Science Foundation of China (51836003, 11872024), National Key Research and Development Project of China No. 2018YFE0127800.

\bibliography{phonon}

\begin{thebibliography}{48}%
\makeatletter
\providecommand \@ifxundefined [1]{%
 \@ifx{#1\undefined}
}%
\providecommand \@ifnum [1]{%
 \ifnum #1\expandafter \@firstoftwo
 \else \expandafter \@secondoftwo
 \fi
}%
\providecommand \@ifx [1]{%
 \ifx #1\expandafter \@firstoftwo
 \else \expandafter \@secondoftwo
 \fi
}%
\providecommand \natexlab [1]{#1}%
\providecommand \enquote  [1]{``#1''}%
\providecommand \bibnamefont  [1]{#1}%
\providecommand \bibfnamefont [1]{#1}%
\providecommand \citenamefont [1]{#1}%
\providecommand \href@noop [0]{\@secondoftwo}%
\providecommand \href [0]{\begingroup \@sanitize@url \@href}%
\providecommand \@href[1]{\@@startlink{#1}\@@href}%
\providecommand \@@href[1]{\endgroup#1\@@endlink}%
\providecommand \@sanitize@url [0]{\catcode `\\12\catcode `\$12\catcode
  `\&12\catcode `\#12\catcode `\^12\catcode `\_12\catcode `\%12\relax}%
\providecommand \@@startlink[1]{}%
\providecommand \@@endlink[0]{}%
\providecommand \url  [0]{\begingroup\@sanitize@url \@url }%
\providecommand \@url [1]{\endgroup\@href {#1}{\urlprefix }}%
\providecommand \urlprefix  [0]{URL }%
\providecommand \Eprint [0]{\href }%
\providecommand \doibase [0]{http://dx.doi.org/}%
\providecommand \selectlanguage [0]{\@gobble}%
\providecommand \bibinfo  [0]{\@secondoftwo}%
\providecommand \bibfield  [0]{\@secondoftwo}%
\providecommand \translation [1]{[#1]}%
\providecommand \BibitemOpen [0]{}%
\providecommand \bibitemStop [0]{}%
\providecommand \bibitemNoStop [0]{.\EOS\space}%
\providecommand \EOS [0]{\spacefactor3000\relax}%
\providecommand \BibitemShut  [1]{\csname bibitem#1\endcsname}%
\let\auto@bib@innerbib\@empty
\bibitem [{\citenamefont {Kaviany}(2008)}]{kaviany_2008}%
  \BibitemOpen
  \bibfield  {author} {\bibinfo {author} {\bibfnamefont {M.}~\bibnamefont
  {Kaviany}},\ }\href {\doibase 10.1017/CBO9780511754586} {\emph {\bibinfo
  {title} {Heat transfer physics}}}\ (\bibinfo  {publisher} {Cambridge
  University Press},\ \bibinfo {year} {2008})\BibitemShut {NoStop}%
\bibitem [{\citenamefont {Chen}(2005)}]{ChenG05Oxford}%
  \BibitemOpen
  \bibfield  {author} {\bibinfo {author} {\bibfnamefont {G.}~\bibnamefont
  {Chen}},\ }\href
  {https://global.oup.com/ushe/product/nanoscale-energy-transport-and-conversion-9780195159424?cc=cn&lang=en&}
  {\emph {\bibinfo {title} {Nanoscale energy transport and conversion: {A}
  parallel treatment of electrons, molecules, phonons, and photons}}}\
  (\bibinfo  {publisher} {Oxford University Press},\ \bibinfo {year}
  {2005})\BibitemShut {NoStop}%
\bibitem [{\citenamefont {Zhang}\ \emph {et~al.}(2020)\citenamefont {Zhang},
  \citenamefont {Ouyang}, \citenamefont {Cheng}, \citenamefont {Chen},
  \citenamefont {Li},\ and\ \citenamefont {Zhang}}]{ZHANG2020}%
  \BibitemOpen
  \bibfield  {author} {\bibinfo {author} {\bibfnamefont {Z.}~\bibnamefont
  {Zhang}}, \bibinfo {author} {\bibfnamefont {Y.}~\bibnamefont {Ouyang}},
  \bibinfo {author} {\bibfnamefont {Y.}~\bibnamefont {Cheng}}, \bibinfo
  {author} {\bibfnamefont {J.}~\bibnamefont {Chen}}, \bibinfo {author}
  {\bibfnamefont {N.}~\bibnamefont {Li}}, \ and\ \bibinfo {author}
  {\bibfnamefont {G.}~\bibnamefont {Zhang}},\ }\href {\doibase
  10.1016/j.physrep.2020.03.001} {\bibfield  {journal} {\bibinfo  {journal}
  {Phys. Rep.}\ }\textbf {\bibinfo {volume} {860}},\ \bibinfo {pages} {1}
  (\bibinfo {year} {2020})}\BibitemShut {NoStop}%
\bibitem [{\citenamefont {Sverdrup}\ \emph {et~al.}(2001)\citenamefont
  {Sverdrup}, \citenamefont {Sinha}, \citenamefont {Asheghi}, \citenamefont
  {Uma},\ and\ \citenamefont {Goodson}}]{ballisticsiliconapl2001}%
  \BibitemOpen
  \bibfield  {author} {\bibinfo {author} {\bibfnamefont {P.~G.}\ \bibnamefont
  {Sverdrup}}, \bibinfo {author} {\bibfnamefont {S.}~\bibnamefont {Sinha}},
  \bibinfo {author} {\bibfnamefont {M.}~\bibnamefont {Asheghi}}, \bibinfo
  {author} {\bibfnamefont {S.}~\bibnamefont {Uma}}, \ and\ \bibinfo {author}
  {\bibfnamefont {K.~E.}\ \bibnamefont {Goodson}},\ }\href {\doibase
  10.1063/1.1371536} {\bibfield  {journal} {\bibinfo  {journal} {Appl. Phys.
  Lett.}\ }\textbf {\bibinfo {volume} {78}},\ \bibinfo {pages} {3331} (\bibinfo
  {year} {2001})}\BibitemShut {NoStop}%
\bibitem [{\citenamefont {Li}\ \emph {et~al.}(2012)\citenamefont {Li},
  \citenamefont {Ren}, \citenamefont {Wang}, \citenamefont {Zhang},
  \citenamefont {H\"anggi},\ and\ \citenamefont {Li}}]{RevModPhysLibaowen}%
  \BibitemOpen
  \bibfield  {author} {\bibinfo {author} {\bibfnamefont {N.}~\bibnamefont
  {Li}}, \bibinfo {author} {\bibfnamefont {J.}~\bibnamefont {Ren}}, \bibinfo
  {author} {\bibfnamefont {L.}~\bibnamefont {Wang}}, \bibinfo {author}
  {\bibfnamefont {G.}~\bibnamefont {Zhang}}, \bibinfo {author} {\bibfnamefont
  {P.}~\bibnamefont {H\"anggi}}, \ and\ \bibinfo {author} {\bibfnamefont
  {B.}~\bibnamefont {Li}},\ }\href {\doibase 10.1103/RevModPhys.84.1045}
  {\bibfield  {journal} {\bibinfo  {journal} {Rev. Mod. Phys.}\ }\textbf
  {\bibinfo {volume} {84}},\ \bibinfo {pages} {1045} (\bibinfo {year}
  {2012})}\BibitemShut {NoStop}%
\bibitem [{\citenamefont {Gu}\ \emph {et~al.}(2018)\citenamefont {Gu},
  \citenamefont {Wei}, \citenamefont {Yin}, \citenamefont {Li},\ and\
  \citenamefont {Yang}}]{RevModPhys.90.041002}%
  \BibitemOpen
  \bibfield  {author} {\bibinfo {author} {\bibfnamefont {X.}~\bibnamefont
  {Gu}}, \bibinfo {author} {\bibfnamefont {Y.}~\bibnamefont {Wei}}, \bibinfo
  {author} {\bibfnamefont {X.}~\bibnamefont {Yin}}, \bibinfo {author}
  {\bibfnamefont {B.}~\bibnamefont {Li}}, \ and\ \bibinfo {author}
  {\bibfnamefont {R.}~\bibnamefont {Yang}},\ }\href {\doibase
  10.1103/RevModPhys.90.041002} {\bibfield  {journal} {\bibinfo  {journal}
  {Rev. Mod. Phys.}\ }\textbf {\bibinfo {volume} {90}},\ \bibinfo {pages}
  {041002} (\bibinfo {year} {2018})}\BibitemShut {NoStop}%
\bibitem [{\citenamefont {Li}\ \emph {et~al.}(2003)\citenamefont {Li},
  \citenamefont {Wu}, \citenamefont {Kim}, \citenamefont {Shi}, \citenamefont
  {Yang},\ and\ \citenamefont {Majumdar}}]{li_thermal_2003}%
  \BibitemOpen
  \bibfield  {author} {\bibinfo {author} {\bibfnamefont {D.}~\bibnamefont
  {Li}}, \bibinfo {author} {\bibfnamefont {Y.}~\bibnamefont {Wu}}, \bibinfo
  {author} {\bibfnamefont {P.}~\bibnamefont {Kim}}, \bibinfo {author}
  {\bibfnamefont {L.}~\bibnamefont {Shi}}, \bibinfo {author} {\bibfnamefont
  {P.}~\bibnamefont {Yang}}, \ and\ \bibinfo {author} {\bibfnamefont
  {A.}~\bibnamefont {Majumdar}},\ }\href {\doibase 10.1063/1.1616981}
  {\bibfield  {journal} {\bibinfo  {journal} {Appl. Phys. Lett.}\ }\textbf
  {\bibinfo {volume} {83}},\ \bibinfo {pages} {2934} (\bibinfo {year}
  {2003})}\BibitemShut {NoStop}%
\bibitem [{\citenamefont {Hsiao}\ \emph {et~al.}(2013)\citenamefont {Hsiao},
  \citenamefont {Chang}, \citenamefont {Liou}, \citenamefont {Chu},
  \citenamefont {Lee},\ and\ \citenamefont {Chang}}]{hsiao_observation_2013}%
  \BibitemOpen
  \bibfield  {author} {\bibinfo {author} {\bibfnamefont {T.-K.}\ \bibnamefont
  {Hsiao}}, \bibinfo {author} {\bibfnamefont {H.-K.}\ \bibnamefont {Chang}},
  \bibinfo {author} {\bibfnamefont {S.-C.}\ \bibnamefont {Liou}}, \bibinfo
  {author} {\bibfnamefont {M.-W.}\ \bibnamefont {Chu}}, \bibinfo {author}
  {\bibfnamefont {S.-C.}\ \bibnamefont {Lee}}, \ and\ \bibinfo {author}
  {\bibfnamefont {C.-W.}\ \bibnamefont {Chang}},\ }\href {\doibase
  10.1038/nnano.2013.121} {\bibfield  {journal} {\bibinfo  {journal} {Nat.
  Nanotechnol.}\ }\textbf {\bibinfo {volume} {8}},\ \bibinfo {pages} {534}
  (\bibinfo {year} {2013})}\BibitemShut {NoStop}%
\bibitem [{\citenamefont {Lee}\ \emph {et~al.}(2017)\citenamefont {Lee},
  \citenamefont {Wu}, \citenamefont {Lou}, \citenamefont {Lee},\ and\
  \citenamefont {Chang}}]{PhysRevLett.118.135901}%
  \BibitemOpen
  \bibfield  {author} {\bibinfo {author} {\bibfnamefont {V.}~\bibnamefont
  {Lee}}, \bibinfo {author} {\bibfnamefont {C.-H.}\ \bibnamefont {Wu}},
  \bibinfo {author} {\bibfnamefont {Z.-X.}\ \bibnamefont {Lou}}, \bibinfo
  {author} {\bibfnamefont {W.-L.}\ \bibnamefont {Lee}}, \ and\ \bibinfo
  {author} {\bibfnamefont {C.-W.}\ \bibnamefont {Chang}},\ }\href {\doibase
  10.1103/PhysRevLett.118.135901} {\bibfield  {journal} {\bibinfo  {journal}
  {Phys. Rev. Lett.}\ }\textbf {\bibinfo {volume} {118}},\ \bibinfo {pages}
  {135901} (\bibinfo {year} {2017})}\BibitemShut {NoStop}%
\bibitem [{\citenamefont {Chang}\ \emph {et~al.}(2008)\citenamefont {Chang},
  \citenamefont {Okawa}, \citenamefont {Garcia}, \citenamefont {Majumdar},\
  and\ \citenamefont {Zettl}}]{chang_breakdown_2008}%
  \BibitemOpen
  \bibfield  {author} {\bibinfo {author} {\bibfnamefont {C.~W.}\ \bibnamefont
  {Chang}}, \bibinfo {author} {\bibfnamefont {D.}~\bibnamefont {Okawa}},
  \bibinfo {author} {\bibfnamefont {H.}~\bibnamefont {Garcia}}, \bibinfo
  {author} {\bibfnamefont {A.}~\bibnamefont {Majumdar}}, \ and\ \bibinfo
  {author} {\bibfnamefont {A.}~\bibnamefont {Zettl}},\ }\href {\doibase
  10.1103/PhysRevLett.101.075903} {\bibfield  {journal} {\bibinfo  {journal}
  {Phys. Rev. Lett.}\ }\textbf {\bibinfo {volume} {101}},\ \bibinfo {pages}
  {075903} (\bibinfo {year} {2008})}\BibitemShut {NoStop}%
\bibitem [{\citenamefont {Xu}\ \emph {et~al.}(2014)\citenamefont {Xu},
  \citenamefont {Pereira}, \citenamefont {Wang}, \citenamefont {Wu},
  \citenamefont {Zhang}, \citenamefont {Zhao}, \citenamefont {Bae},
  \citenamefont {Bui}, \citenamefont {Xie}, \citenamefont {Thong},
  \citenamefont {Hong}, \citenamefont {Loh}, \citenamefont {Donadio},
  \citenamefont {Li},\ and\ \citenamefont
  {{\"O}zyilmaz}}]{xu_length-dependent_2014}%
  \BibitemOpen
  \bibfield  {author} {\bibinfo {author} {\bibfnamefont {X.}~\bibnamefont
  {Xu}}, \bibinfo {author} {\bibfnamefont {L.~F.~C.}\ \bibnamefont {Pereira}},
  \bibinfo {author} {\bibfnamefont {Y.}~\bibnamefont {Wang}}, \bibinfo {author}
  {\bibfnamefont {J.}~\bibnamefont {Wu}}, \bibinfo {author} {\bibfnamefont
  {K.}~\bibnamefont {Zhang}}, \bibinfo {author} {\bibfnamefont
  {X.}~\bibnamefont {Zhao}}, \bibinfo {author} {\bibfnamefont {S.}~\bibnamefont
  {Bae}}, \bibinfo {author} {\bibfnamefont {C.~T.}\ \bibnamefont {Bui}},
  \bibinfo {author} {\bibfnamefont {R.}~\bibnamefont {Xie}}, \bibinfo {author}
  {\bibfnamefont {J.~T.~L.}\ \bibnamefont {Thong}}, \bibinfo {author}
  {\bibfnamefont {B.~H.}\ \bibnamefont {Hong}}, \bibinfo {author}
  {\bibfnamefont {K.~P.}\ \bibnamefont {Loh}}, \bibinfo {author} {\bibfnamefont
  {D.}~\bibnamefont {Donadio}}, \bibinfo {author} {\bibfnamefont
  {B.}~\bibnamefont {Li}}, \ and\ \bibinfo {author} {\bibfnamefont
  {B.}~\bibnamefont {{\"O}zyilmaz}},\ }\href {\doibase 10.1038/ncomms4689}
  {\bibfield  {journal} {\bibinfo  {journal} {Nat. Commun.}\ }\textbf {\bibinfo
  {volume} {5}},\ \bibinfo {pages} {3689} (\bibinfo {year} {2014})}\BibitemShut
  {NoStop}%
\bibitem [{\citenamefont {Mahan}\ and\ \citenamefont
  {Claro}(1988)}]{PhysRevB.38.1963}%
  \BibitemOpen
  \bibfield  {author} {\bibinfo {author} {\bibfnamefont {G.~D.}\ \bibnamefont
  {Mahan}}\ and\ \bibinfo {author} {\bibfnamefont {F.}~\bibnamefont {Claro}},\
  }\href {\doibase 10.1103/PhysRevB.38.1963} {\bibfield  {journal} {\bibinfo
  {journal} {Phys. Rev. B}\ }\textbf {\bibinfo {volume} {38}},\ \bibinfo
  {pages} {1963} (\bibinfo {year} {1988})}\BibitemShut {NoStop}%
\bibitem [{\citenamefont {Chen}(1996)}]{chen1996}%
  \BibitemOpen
  \bibfield  {author} {\bibinfo {author} {\bibfnamefont {G.}~\bibnamefont
  {Chen}},\ }\href {\doibase 10.1115/1.2822665} {\bibfield  {journal} {\bibinfo
   {journal} {J. Heat Transfer}\ }\textbf {\bibinfo {volume} {118}},\ \bibinfo
  {pages} {539} (\bibinfo {year} {1996})}\BibitemShut {NoStop}%
\bibitem [{\citenamefont {Regner}\ \emph {et~al.}(2014)\citenamefont {Regner},
  \citenamefont {McGaughey},\ and\ \citenamefont {Malen}}]{PhysRevB.90.064302}%
  \BibitemOpen
  \bibfield  {author} {\bibinfo {author} {\bibfnamefont {K.~T.}\ \bibnamefont
  {Regner}}, \bibinfo {author} {\bibfnamefont {A.~J.~H.}\ \bibnamefont
  {McGaughey}}, \ and\ \bibinfo {author} {\bibfnamefont {J.~A.}\ \bibnamefont
  {Malen}},\ }\href {\doibase 10.1103/PhysRevB.90.064302} {\bibfield  {journal}
  {\bibinfo  {journal} {Phys. Rev. B}\ }\textbf {\bibinfo {volume} {90}},\
  \bibinfo {pages} {064302} (\bibinfo {year} {2014})}\BibitemShut {NoStop}%
\bibitem [{\citenamefont {Majumdar}(1993)}]{MajumdarA93Film}%
  \BibitemOpen
  \bibfield  {author} {\bibinfo {author} {\bibfnamefont {A.}~\bibnamefont
  {Majumdar}},\ }\href {\doibase 10.1115/1.2910673} {\bibfield  {journal}
  {\bibinfo  {journal} {J. Heat Transfer}\ }\textbf {\bibinfo {volume} {115}},\
  \bibinfo {pages} {7} (\bibinfo {year} {1993})}\BibitemShut {NoStop}%
\bibitem [{\citenamefont {Volz}\ and\ \citenamefont {Chen}(1999)}]{volz1999}%
  \BibitemOpen
  \bibfield  {author} {\bibinfo {author} {\bibfnamefont {S.~G.}\ \bibnamefont
  {Volz}}\ and\ \bibinfo {author} {\bibfnamefont {G.}~\bibnamefont {Chen}},\
  }\href {\doibase 10.1063/1.124914} {\bibfield  {journal} {\bibinfo  {journal}
  {Appl. Phys. Lett.}\ }\textbf {\bibinfo {volume} {75}},\ \bibinfo {pages}
  {2056} (\bibinfo {year} {1999})}\BibitemShut {NoStop}%
\bibitem [{\citenamefont {Prosen}\ and\ \citenamefont
  {Campbell}(2000)}]{PhysRevLett.84.2857}%
  \BibitemOpen
  \bibfield  {author} {\bibinfo {author} {\bibfnamefont {T.~c.~v.}\
  \bibnamefont {Prosen}}\ and\ \bibinfo {author} {\bibfnamefont {D.~K.}\
  \bibnamefont {Campbell}},\ }\href {\doibase 10.1103/PhysRevLett.84.2857}
  {\bibfield  {journal} {\bibinfo  {journal} {Phys. Rev. Lett.}\ }\textbf
  {\bibinfo {volume} {84}},\ \bibinfo {pages} {2857} (\bibinfo {year}
  {2000})}\BibitemShut {NoStop}%
\bibitem [{\citenamefont {Narayan}\ and\ \citenamefont
  {Ramaswamy}(2002)}]{narayan_anomalous_2002}%
  \BibitemOpen
  \bibfield  {author} {\bibinfo {author} {\bibfnamefont {O.}~\bibnamefont
  {Narayan}}\ and\ \bibinfo {author} {\bibfnamefont {S.}~\bibnamefont
  {Ramaswamy}},\ }\href {\doibase 10.1103/PhysRevLett.89.200601} {\bibfield
  {journal} {\bibinfo  {journal} {Phys. Rev. Lett.}\ }\textbf {\bibinfo
  {volume} {89}},\ \bibinfo {pages} {200601} (\bibinfo {year}
  {2002})}\BibitemShut {NoStop}%
\bibitem [{\citenamefont {Lepri}\ \emph {et~al.}(2003)\citenamefont {Lepri},
  \citenamefont {Livi},\ and\ \citenamefont {Politi}}]{lepri_thermal_2003}%
  \BibitemOpen
  \bibfield  {author} {\bibinfo {author} {\bibfnamefont {S.}~\bibnamefont
  {Lepri}}, \bibinfo {author} {\bibfnamefont {R.}~\bibnamefont {Livi}}, \ and\
  \bibinfo {author} {\bibfnamefont {A.}~\bibnamefont {Politi}},\ }\href
  {\doibase https://doi.org/10.1016/S0370-1573(02)00558-6} {\bibfield
  {journal} {\bibinfo  {journal} {Phys. Rep.}\ }\textbf {\bibinfo {volume}
  {377}},\ \bibinfo {pages} {1 } (\bibinfo {year} {2003})}\BibitemShut
  {NoStop}%
\bibitem [{\citenamefont {Dhar}(2008)}]{Dhar2008advphys}%
  \BibitemOpen
  \bibfield  {author} {\bibinfo {author} {\bibfnamefont {A.}~\bibnamefont
  {Dhar}},\ }\href {\doibase 10.1080/00018730802538522} {\bibfield  {journal}
  {\bibinfo  {journal} {Adv. Phys.}\ }\textbf {\bibinfo {volume} {57}},\
  \bibinfo {pages} {457} (\bibinfo {year} {2008})}\BibitemShut {NoStop}%
\bibitem [{\citenamefont {Yang}\ and\ \citenamefont
  {Chen}(2003)}]{PhysRevB.67.195311}%
  \BibitemOpen
  \bibfield  {author} {\bibinfo {author} {\bibfnamefont {B.}~\bibnamefont
  {Yang}}\ and\ \bibinfo {author} {\bibfnamefont {G.}~\bibnamefont {Chen}},\
  }\href {\doibase 10.1103/PhysRevB.67.195311} {\bibfield  {journal} {\bibinfo
  {journal} {Phys. Rev. B}\ }\textbf {\bibinfo {volume} {67}},\ \bibinfo
  {pages} {195311} (\bibinfo {year} {2003})}\BibitemShut {NoStop}%
\bibitem [{\citenamefont {Ma}\ \emph {et~al.}(2019)\citenamefont {Ma},
  \citenamefont {Arora}, \citenamefont {Deng}, \citenamefont {Xie},
  \citenamefont {Shiomi},\ and\ \citenamefont {Yang}}]{ma2019b}%
  \BibitemOpen
  \bibfield  {author} {\bibinfo {author} {\bibfnamefont {D.}~\bibnamefont
  {Ma}}, \bibinfo {author} {\bibfnamefont {A.}~\bibnamefont {Arora}}, \bibinfo
  {author} {\bibfnamefont {S.}~\bibnamefont {Deng}}, \bibinfo {author}
  {\bibfnamefont {G.}~\bibnamefont {Xie}}, \bibinfo {author} {\bibfnamefont
  {J.}~\bibnamefont {Shiomi}}, \ and\ \bibinfo {author} {\bibfnamefont
  {N.}~\bibnamefont {Yang}},\ }\href {\doibase 10.1016/j.mtphys.2019.01.002}
  {\bibfield  {journal} {\bibinfo  {journal} {Mater. Today Phys.}\ }\textbf
  {\bibinfo {volume} {8}},\ \bibinfo {pages} {56} (\bibinfo {year}
  {2019})}\BibitemShut {NoStop}%
\bibitem [{\citenamefont {Guyer}\ and\ \citenamefont
  {Krumhansl}(1966{\natexlab{a}})}]{PhysRev_GK}%
  \BibitemOpen
  \bibfield  {author} {\bibinfo {author} {\bibfnamefont {R.~A.}\ \bibnamefont
  {Guyer}}\ and\ \bibinfo {author} {\bibfnamefont {J.~A.}\ \bibnamefont
  {Krumhansl}},\ }\href {\doibase 10.1103/PhysRev.148.778} {\bibfield
  {journal} {\bibinfo  {journal} {Phys. Rev.}\ }\textbf {\bibinfo {volume}
  {148}},\ \bibinfo {pages} {778} (\bibinfo {year}
  {1966}{\natexlab{a}})}\BibitemShut {NoStop}%
\bibitem [{\citenamefont {Guyer}\ and\ \citenamefont
  {Krumhansl}(1966{\natexlab{b}})}]{PhysRev.148.766}%
  \BibitemOpen
  \bibfield  {author} {\bibinfo {author} {\bibfnamefont {R.~A.}\ \bibnamefont
  {Guyer}}\ and\ \bibinfo {author} {\bibfnamefont {J.~A.}\ \bibnamefont
  {Krumhansl}},\ }\href {\doibase 10.1103/PhysRev.148.766} {\bibfield
  {journal} {\bibinfo  {journal} {Phys. Rev.}\ }\textbf {\bibinfo {volume}
  {148}},\ \bibinfo {pages} {766} (\bibinfo {year}
  {1966}{\natexlab{b}})}\BibitemShut {NoStop}%
\bibitem [{\citenamefont {Lee}\ and\ \citenamefont
  {Li}(2020)}]{leesangyeopch1}%
  \BibitemOpen
  \bibfield  {author} {\bibinfo {author} {\bibfnamefont {S.}~\bibnamefont
  {Lee}}\ and\ \bibinfo {author} {\bibfnamefont {X.}~\bibnamefont {Li}},\ }in\
  \href {\doibase 10.1088/978-0-7503-1738-2ch1} {\emph {\bibinfo {booktitle}
  {Nanoscale Energy Transport}}},\ \bibinfo {series and number} {2053-2563}\
  (\bibinfo  {publisher} {IOP Publishing},\ \bibinfo {year} {2020})\ pp.\
  \bibinfo {pages} {1--1 to 1--26}\BibitemShut {NoStop}%
\bibitem [{\citenamefont {Lee}\ \emph {et~al.}(2015)\citenamefont {Lee},
  \citenamefont {Broido}, \citenamefont {Esfarjani},\ and\ \citenamefont
  {Chen}}]{lee_hydrodynamic_2015}%
  \BibitemOpen
  \bibfield  {author} {\bibinfo {author} {\bibfnamefont {S.}~\bibnamefont
  {Lee}}, \bibinfo {author} {\bibfnamefont {D.}~\bibnamefont {Broido}},
  \bibinfo {author} {\bibfnamefont {K.}~\bibnamefont {Esfarjani}}, \ and\
  \bibinfo {author} {\bibfnamefont {G.}~\bibnamefont {Chen}},\ }\href {\doibase
  10.1038/ncomms7290} {\bibfield  {journal} {\bibinfo  {journal} {Nat.
  Commun.}\ }\textbf {\bibinfo {volume} {6}},\ \bibinfo {pages} {6290}
  (\bibinfo {year} {2015})}\BibitemShut {NoStop}%
\bibitem [{\citenamefont {Cepellotti}\ \emph {et~al.}(2015)\citenamefont
  {Cepellotti}, \citenamefont {Fugallo}, \citenamefont {Paulatto},
  \citenamefont {Lazzeri}, \citenamefont {Mauri},\ and\ \citenamefont
  {Marzari}}]{cepellotti_phonon_2015}%
  \BibitemOpen
  \bibfield  {author} {\bibinfo {author} {\bibfnamefont {A.}~\bibnamefont
  {Cepellotti}}, \bibinfo {author} {\bibfnamefont {G.}~\bibnamefont {Fugallo}},
  \bibinfo {author} {\bibfnamefont {L.}~\bibnamefont {Paulatto}}, \bibinfo
  {author} {\bibfnamefont {M.}~\bibnamefont {Lazzeri}}, \bibinfo {author}
  {\bibfnamefont {F.}~\bibnamefont {Mauri}}, \ and\ \bibinfo {author}
  {\bibfnamefont {N.}~\bibnamefont {Marzari}},\ }\href {\doibase
  10.1038/ncomms7400} {\bibfield  {journal} {\bibinfo  {journal} {Nat.
  Commun.}\ }\textbf {\bibinfo {volume} {6}},\ \bibinfo {pages} {6400}
  (\bibinfo {year} {2015})}\BibitemShut {NoStop}%
\bibitem [{\citenamefont {Guo}\ and\ \citenamefont
  {Wang}(2015)}]{WangMr15application}%
  \BibitemOpen
  \bibfield  {author} {\bibinfo {author} {\bibfnamefont {Y.}~\bibnamefont
  {Guo}}\ and\ \bibinfo {author} {\bibfnamefont {M.}~\bibnamefont {Wang}},\
  }\href {\doibase 10.1016/j.physrep.2015.07.003} {\bibfield  {journal}
  {\bibinfo  {journal} {Phys. Rep.}\ }\textbf {\bibinfo {volume} {595}},\
  \bibinfo {pages} {1 } (\bibinfo {year} {2015})}\BibitemShut {NoStop}%
\bibitem [{\citenamefont {Huberman}\ \emph {et~al.}(2019)\citenamefont
  {Huberman}, \citenamefont {Duncan}, \citenamefont {Chen}, \citenamefont
  {Song}, \citenamefont {Chiloyan}, \citenamefont {Ding}, \citenamefont
  {Maznev}, \citenamefont {Chen},\ and\ \citenamefont
  {Nelson}}]{huberman_observation_2019}%
  \BibitemOpen
  \bibfield  {author} {\bibinfo {author} {\bibfnamefont {S.}~\bibnamefont
  {Huberman}}, \bibinfo {author} {\bibfnamefont {R.~A.}\ \bibnamefont
  {Duncan}}, \bibinfo {author} {\bibfnamefont {K.}~\bibnamefont {Chen}},
  \bibinfo {author} {\bibfnamefont {B.}~\bibnamefont {Song}}, \bibinfo {author}
  {\bibfnamefont {V.}~\bibnamefont {Chiloyan}}, \bibinfo {author}
  {\bibfnamefont {Z.}~\bibnamefont {Ding}}, \bibinfo {author} {\bibfnamefont
  {A.~A.}\ \bibnamefont {Maznev}}, \bibinfo {author} {\bibfnamefont
  {G.}~\bibnamefont {Chen}}, \ and\ \bibinfo {author} {\bibfnamefont {K.~A.}\
  \bibnamefont {Nelson}},\ }\href {\doibase 10.1126/science.aav3548} {\bibfield
   {journal} {\bibinfo  {journal} {Science}\ }\textbf {\bibinfo {volume}
  {364}},\ \bibinfo {pages} {375} (\bibinfo {year} {2019})}\BibitemShut
  {NoStop}%
\bibitem [{\citenamefont {Hua}\ and\ \citenamefont
  {Minnich}(2018)}]{PhysRevB.97.014307}%
  \BibitemOpen
  \bibfield  {author} {\bibinfo {author} {\bibfnamefont {C.}~\bibnamefont
  {Hua}}\ and\ \bibinfo {author} {\bibfnamefont {A.~J.}\ \bibnamefont
  {Minnich}},\ }\href {\doibase 10.1103/PhysRevB.97.014307} {\bibfield
  {journal} {\bibinfo  {journal} {Phys. Rev. B}\ }\textbf {\bibinfo {volume}
  {97}},\ \bibinfo {pages} {014307} (\bibinfo {year} {2018})}\BibitemShut
  {NoStop}%
\bibitem [{\citenamefont {Chen}\ \emph {et~al.}(2018)\citenamefont {Chen},
  \citenamefont {Hua}, \citenamefont {Zhang}, \citenamefont {Ravichandran},\
  and\ \citenamefont {Minnich}}]{PhysRevApplied.10.054068}%
  \BibitemOpen
  \bibfield  {author} {\bibinfo {author} {\bibfnamefont {X.}~\bibnamefont
  {Chen}}, \bibinfo {author} {\bibfnamefont {C.}~\bibnamefont {Hua}}, \bibinfo
  {author} {\bibfnamefont {H.}~\bibnamefont {Zhang}}, \bibinfo {author}
  {\bibfnamefont {N.~K.}\ \bibnamefont {Ravichandran}}, \ and\ \bibinfo
  {author} {\bibfnamefont {A.~J.}\ \bibnamefont {Minnich}},\ }\href {\doibase
  10.1103/PhysRevApplied.10.054068} {\bibfield  {journal} {\bibinfo  {journal}
  {Phys. Rev. Applied}\ }\textbf {\bibinfo {volume} {10}},\ \bibinfo {pages}
  {054068} (\bibinfo {year} {2018})}\BibitemShut {NoStop}%
\bibitem [{\citenamefont {Minnich}\ \emph {et~al.}(2011)\citenamefont
  {Minnich}, \citenamefont {Johnson}, \citenamefont {Schmidt}, \citenamefont
  {Esfarjani}, \citenamefont {Dresselhaus}, \citenamefont {Nelson},\ and\
  \citenamefont {Chen}}]{PhysRevLett.107.095901}%
  \BibitemOpen
  \bibfield  {author} {\bibinfo {author} {\bibfnamefont {A.~J.}\ \bibnamefont
  {Minnich}}, \bibinfo {author} {\bibfnamefont {J.~A.}\ \bibnamefont
  {Johnson}}, \bibinfo {author} {\bibfnamefont {A.~J.}\ \bibnamefont
  {Schmidt}}, \bibinfo {author} {\bibfnamefont {K.}~\bibnamefont {Esfarjani}},
  \bibinfo {author} {\bibfnamefont {M.~S.}\ \bibnamefont {Dresselhaus}},
  \bibinfo {author} {\bibfnamefont {K.~A.}\ \bibnamefont {Nelson}}, \ and\
  \bibinfo {author} {\bibfnamefont {G.}~\bibnamefont {Chen}},\ }\href {\doibase
  10.1103/PhysRevLett.107.095901} {\bibfield  {journal} {\bibinfo  {journal}
  {Phys. Rev. Lett.}\ }\textbf {\bibinfo {volume} {107}},\ \bibinfo {pages}
  {095901} (\bibinfo {year} {2011})}\BibitemShut {NoStop}%
\bibitem [{\citenamefont {Hu}\ \emph {et~al.}(2015)\citenamefont {Hu},
  \citenamefont {Zeng}, \citenamefont {Minnich}, \citenamefont {Dresselhaus},\
  and\ \citenamefont {Chen}}]{hu_spectral_2015}%
  \BibitemOpen
  \bibfield  {author} {\bibinfo {author} {\bibfnamefont {Y.}~\bibnamefont
  {Hu}}, \bibinfo {author} {\bibfnamefont {L.}~\bibnamefont {Zeng}}, \bibinfo
  {author} {\bibfnamefont {A.~J.}\ \bibnamefont {Minnich}}, \bibinfo {author}
  {\bibfnamefont {M.~S.}\ \bibnamefont {Dresselhaus}}, \ and\ \bibinfo {author}
  {\bibfnamefont {G.}~\bibnamefont {Chen}},\ }\href {\doibase
  10.1038/nnano.2015.109} {\bibfield  {journal} {\bibinfo  {journal} {Nat.
  Nanotechnol.}\ }\textbf {\bibinfo {volume} {10}},\ \bibinfo {pages} {701}
  (\bibinfo {year} {2015})}\BibitemShut {NoStop}%
\bibitem [{\citenamefont {Siemens}\ \emph {et~al.}(2010)\citenamefont
  {Siemens}, \citenamefont {Li}, \citenamefont {Yang}, \citenamefont {Nelson},
  \citenamefont {Anderson}, \citenamefont {Murnane},\ and\ \citenamefont
  {Kapteyn}}]{siemens_quasi-ballistic_2010}%
  \BibitemOpen
  \bibfield  {author} {\bibinfo {author} {\bibfnamefont {M.~E.}\ \bibnamefont
  {Siemens}}, \bibinfo {author} {\bibfnamefont {Q.}~\bibnamefont {Li}},
  \bibinfo {author} {\bibfnamefont {R.}~\bibnamefont {Yang}}, \bibinfo {author}
  {\bibfnamefont {K.~A.}\ \bibnamefont {Nelson}}, \bibinfo {author}
  {\bibfnamefont {E.~H.}\ \bibnamefont {Anderson}}, \bibinfo {author}
  {\bibfnamefont {M.~M.}\ \bibnamefont {Murnane}}, \ and\ \bibinfo {author}
  {\bibfnamefont {H.~C.}\ \bibnamefont {Kapteyn}},\ }\href {\doibase
  10.1038/nmat2568} {\bibfield  {journal} {\bibinfo  {journal} {Nat. Mater.}\
  }\textbf {\bibinfo {volume} {9}},\ \bibinfo {pages} {26} (\bibinfo {year}
  {2010})}\BibitemShut {NoStop}%
\bibitem [{\citenamefont {Yang}\ \emph {et~al.}(2015)\citenamefont {Yang},
  \citenamefont {Hu}, \citenamefont {Ma}, \citenamefont {Lu},\ and\
  \citenamefont {Li}}]{yang_nanoscale_2015}%
  \BibitemOpen
  \bibfield  {author} {\bibinfo {author} {\bibfnamefont {N.}~\bibnamefont
  {Yang}}, \bibinfo {author} {\bibfnamefont {S.}~\bibnamefont {Hu}}, \bibinfo
  {author} {\bibfnamefont {D.}~\bibnamefont {Ma}}, \bibinfo {author}
  {\bibfnamefont {T.}~\bibnamefont {Lu}}, \ and\ \bibinfo {author}
  {\bibfnamefont {B.}~\bibnamefont {Li}},\ }\href {\doibase 10.1038/srep14878}
  {\bibfield  {journal} {\bibinfo  {journal} {Sci. Rep.}\ }\textbf {\bibinfo
  {volume} {5}},\ \bibinfo {pages} {14878} (\bibinfo {year}
  {2015})}\BibitemShut {NoStop}%
\bibitem [{\citenamefont {Ma}\ \emph {et~al.}(2017)\citenamefont {Ma},
  \citenamefont {Ding}, \citenamefont {Wang}, \citenamefont {Yang},\ and\
  \citenamefont {Zhang}}]{ma_unexpected_2017}%
  \BibitemOpen
  \bibfield  {author} {\bibinfo {author} {\bibfnamefont {D.}~\bibnamefont
  {Ma}}, \bibinfo {author} {\bibfnamefont {H.}~\bibnamefont {Ding}}, \bibinfo
  {author} {\bibfnamefont {X.}~\bibnamefont {Wang}}, \bibinfo {author}
  {\bibfnamefont {N.}~\bibnamefont {Yang}}, \ and\ \bibinfo {author}
  {\bibfnamefont {X.}~\bibnamefont {Zhang}},\ }\href {\doibase
  https://doi.org/10.1016/j.ijheatmasstransfer.2016.12.092} {\bibfield
  {journal} {\bibinfo  {journal} {Int. J. Heat Mass Transfer}\ }\textbf
  {\bibinfo {volume} {108}},\ \bibinfo {pages} {940 } (\bibinfo {year}
  {2017})}\BibitemShut {NoStop}%
\bibitem [{\citenamefont {Callaway}(1959)}]{PhysRev_callaway}%
  \BibitemOpen
  \bibfield  {author} {\bibinfo {author} {\bibfnamefont {J.}~\bibnamefont
  {Callaway}},\ }\href {\doibase 10.1103/PhysRev.113.1046} {\bibfield
  {journal} {\bibinfo  {journal} {Phys. Rev.}\ }\textbf {\bibinfo {volume}
  {113}},\ \bibinfo {pages} {1046} (\bibinfo {year} {1959})}\BibitemShut
  {NoStop}%
\bibitem [{\citenamefont {Guo}\ and\ \citenamefont
  {Wang}(2017)}]{wangmr17callaway}%
  \BibitemOpen
  \bibfield  {author} {\bibinfo {author} {\bibfnamefont {Y.}~\bibnamefont
  {Guo}}\ and\ \bibinfo {author} {\bibfnamefont {M.}~\bibnamefont {Wang}},\
  }\href {\doibase 10.1103/PhysRevB.96.134312} {\bibfield  {journal} {\bibinfo
  {journal} {Phys. Rev. B}\ }\textbf {\bibinfo {volume} {96}},\ \bibinfo
  {pages} {134312} (\bibinfo {year} {2017})}\BibitemShut {NoStop}%
\bibitem [{\citenamefont {Allen}(2018)}]{PhysRevB.97.134307}%
  \BibitemOpen
  \bibfield  {author} {\bibinfo {author} {\bibfnamefont {P.~B.}\ \bibnamefont
  {Allen}},\ }\href {\doibase 10.1103/PhysRevB.97.134307} {\bibfield  {journal}
  {\bibinfo  {journal} {Phys. Rev. B}\ }\textbf {\bibinfo {volume} {97}},\
  \bibinfo {pages} {134307} (\bibinfo {year} {2018})}\BibitemShut {NoStop}%
\bibitem [{\citenamefont {Luo}\ \emph {et~al.}(2019)\citenamefont {Luo},
  \citenamefont {Guo}, \citenamefont {Wang},\ and\ \citenamefont
  {Yi}}]{luo2019}%
  \BibitemOpen
  \bibfield  {author} {\bibinfo {author} {\bibfnamefont {X.-P.}\ \bibnamefont
  {Luo}}, \bibinfo {author} {\bibfnamefont {Y.-Y.}\ \bibnamefont {Guo}},
  \bibinfo {author} {\bibfnamefont {M.-R.}\ \bibnamefont {Wang}}, \ and\
  \bibinfo {author} {\bibfnamefont {H.-L.}\ \bibnamefont {Yi}},\ }\href
  {\doibase 10.1103/PhysRevB.100.155401} {\bibfield  {journal} {\bibinfo
  {journal} {Phys. Rev. B}\ }\textbf {\bibinfo {volume} {100}},\ \bibinfo
  {pages} {155401} (\bibinfo {year} {2019})}\BibitemShut {NoStop}%
\bibitem [{\citenamefont {Li}\ and\ \citenamefont {Cao}(2019)}]{li2019}%
  \BibitemOpen
  \bibfield  {author} {\bibinfo {author} {\bibfnamefont {H.-L.}\ \bibnamefont
  {Li}}\ and\ \bibinfo {author} {\bibfnamefont {B.-Y.}\ \bibnamefont {Cao}},\
  }\href {\doibase 10.1080/15567265.2018.1520763} {\bibfield  {journal}
  {\bibinfo  {journal} {Nanosc. Microsc. Therm.}\ }\textbf {\bibinfo {volume}
  {23}},\ \bibinfo {pages} {10} (\bibinfo {year} {2019})}\BibitemShut {NoStop}%
\bibitem [{\citenamefont {Zhang}\ \emph {et~al.}(2019)\citenamefont {Zhang},
  \citenamefont {Guo},\ and\ \citenamefont {Chen}}]{ZHANG20191366}%
  \BibitemOpen
  \bibfield  {author} {\bibinfo {author} {\bibfnamefont {C.}~\bibnamefont
  {Zhang}}, \bibinfo {author} {\bibfnamefont {Z.}~\bibnamefont {Guo}}, \ and\
  \bibinfo {author} {\bibfnamefont {S.}~\bibnamefont {Chen}},\ }\href {\doibase
  10.1016/j.ijheatmasstransfer.2018.10.141} {\bibfield  {journal} {\bibinfo
  {journal} {Int. J. Heat Mass Transfer}\ }\textbf {\bibinfo {volume} {130}},\
  \bibinfo {pages} {1366} (\bibinfo {year} {2019})}\BibitemShut {NoStop}%
\bibitem [{\citenamefont {Olfe}(1968)}]{olfe1968}%
  \BibitemOpen
  \bibfield  {author} {\bibinfo {author} {\bibfnamefont {D.~B.}\ \bibnamefont
  {Olfe}},\ }\href {\doibase 10.1016/0022-4073(68)90094-0} {\bibfield
  {journal} {\bibinfo  {journal} {J. Quant. Spectrosc. Ra.}\ }\textbf {\bibinfo
  {volume} {8}},\ \bibinfo {pages} {899} (\bibinfo {year} {1968})}\BibitemShut
  {NoStop}%
\bibitem [{\citenamefont {Yang}\ \emph {et~al.}(2019)\citenamefont {Yang},
  \citenamefont {Yue},\ and\ \citenamefont {Liao}}]{Nanalytical}%
  \BibitemOpen
  \bibfield  {author} {\bibinfo {author} {\bibfnamefont {R.}~\bibnamefont
  {Yang}}, \bibinfo {author} {\bibfnamefont {S.}~\bibnamefont {Yue}}, \ and\
  \bibinfo {author} {\bibfnamefont {B.}~\bibnamefont {Liao}},\ }\href {\doibase
  10.1080/15567265.2018.1551449} {\bibfield  {journal} {\bibinfo  {journal}
  {Nanosc. Microsc. Therm.}\ }\textbf {\bibinfo {volume} {23}},\ \bibinfo
  {pages} {25} (\bibinfo {year} {2019})}\BibitemShut {NoStop}%
\bibitem [{\citenamefont {Li}\ and\ \citenamefont
  {Lee}(2019)}]{PhysRevB.99.085202}%
  \BibitemOpen
  \bibfield  {author} {\bibinfo {author} {\bibfnamefont {X.}~\bibnamefont
  {Li}}\ and\ \bibinfo {author} {\bibfnamefont {S.}~\bibnamefont {Lee}},\
  }\href {\doibase 10.1103/PhysRevB.99.085202} {\bibfield  {journal} {\bibinfo
  {journal} {Phys. Rev. B}\ }\textbf {\bibinfo {volume} {99}},\ \bibinfo
  {pages} {085202} (\bibinfo {year} {2019})}\BibitemShut {NoStop}%
\bibitem [{\citenamefont {Barenblatt}(1987)}]{barenblatt1987dimensional}%
  \BibitemOpen
  \bibfield  {author} {\bibinfo {author} {\bibfnamefont {G.~I.}\ \bibnamefont
  {Barenblatt}},\ }\href@noop {} {\emph {\bibinfo {title} {Dimensional
  analysis}}}\ (\bibinfo  {publisher} {CRC Press},\ \bibinfo {year}
  {1987})\BibitemShut {NoStop}%
\bibitem [{\citenamefont {Balandin}\ \emph {et~al.}(2008)\citenamefont
  {Balandin}, \citenamefont {Ghosh}, \citenamefont {Bao}, \citenamefont
  {Calizo}, \citenamefont {Teweldebrhan}, \citenamefont {Miao},\ and\
  \citenamefont {Lau}}]{balandin_superior_2008}%
  \BibitemOpen
  \bibfield  {author} {\bibinfo {author} {\bibfnamefont {A.~A.}\ \bibnamefont
  {Balandin}}, \bibinfo {author} {\bibfnamefont {S.}~\bibnamefont {Ghosh}},
  \bibinfo {author} {\bibfnamefont {W.}~\bibnamefont {Bao}}, \bibinfo {author}
  {\bibfnamefont {I.}~\bibnamefont {Calizo}}, \bibinfo {author} {\bibfnamefont
  {D.}~\bibnamefont {Teweldebrhan}}, \bibinfo {author} {\bibfnamefont
  {F.}~\bibnamefont {Miao}}, \ and\ \bibinfo {author} {\bibfnamefont {C.~N.}\
  \bibnamefont {Lau}},\ }\href {\doibase 10.1021/nl0731872} {\bibfield
  {journal} {\bibinfo  {journal} {Nano Lett.}\ }\textbf {\bibinfo {volume}
  {8}},\ \bibinfo {pages} {902} (\bibinfo {year} {2008})}\BibitemShut {NoStop}%
\bibitem [{\citenamefont {Bae}\ \emph {et~al.}(2013)\citenamefont {Bae},
  \citenamefont {Li}, \citenamefont {Aksamija}, \citenamefont {Martin},
  \citenamefont {Xiong}, \citenamefont {Ong}, \citenamefont {Knezevic},\ and\
  \citenamefont {Pop}}]{bae2013ballistic}%
  \BibitemOpen
  \bibfield  {author} {\bibinfo {author} {\bibfnamefont {M.-H.}\ \bibnamefont
  {Bae}}, \bibinfo {author} {\bibfnamefont {Z.}~\bibnamefont {Li}}, \bibinfo
  {author} {\bibfnamefont {Z.}~\bibnamefont {Aksamija}}, \bibinfo {author}
  {\bibfnamefont {P.~N.}\ \bibnamefont {Martin}}, \bibinfo {author}
  {\bibfnamefont {F.}~\bibnamefont {Xiong}}, \bibinfo {author} {\bibfnamefont
  {Z.-Y.}\ \bibnamefont {Ong}}, \bibinfo {author} {\bibfnamefont
  {I.}~\bibnamefont {Knezevic}}, \ and\ \bibinfo {author} {\bibfnamefont
  {E.}~\bibnamefont {Pop}},\ }\href {\doibase 10.1038/ncomms2755} {\bibfield
  {journal} {\bibinfo  {journal} {Nat. Commun.}\ }\textbf {\bibinfo {volume}
  {4}},\ \bibinfo {pages} {1734} (\bibinfo {year} {2013})}\BibitemShut
  {NoStop}%
\end{thebibliography}%


\begin{thebibliography}{23}%
\makeatletter
\providecommand \@ifxundefined [1]{%
 \@ifx{#1\undefined}
}%
\providecommand \@ifnum [1]{%
 \ifnum #1\expandafter \@firstoftwo
 \else \expandafter \@secondoftwo
 \fi
}%
\providecommand \@ifx [1]{%
 \ifx #1\expandafter \@firstoftwo
 \else \expandafter \@secondoftwo
 \fi
}%
\providecommand \natexlab [1]{#1}%
\providecommand \enquote  [1]{``#1''}%
\providecommand \bibnamefont  [1]{#1}%
\providecommand \bibfnamefont [1]{#1}%
\providecommand \citenamefont [1]{#1}%
\providecommand \href@noop [0]{\@secondoftwo}%
\providecommand \href [0]{\begingroup \@sanitize@url \@href}%
\providecommand \@href[1]{\@@startlink{#1}\@@href}%
\providecommand \@@href[1]{\endgroup#1\@@endlink}%
\providecommand \@sanitize@url [0]{\catcode `\\12\catcode `\$12\catcode
  `\&12\catcode `\#12\catcode `\^12\catcode `\_12\catcode `\%12\relax}%
\providecommand \@@startlink[1]{}%
\providecommand \@@endlink[0]{}%
\providecommand \url  [0]{\begingroup\@sanitize@url \@url }%
\providecommand \@url [1]{\endgroup\@href {#1}{\urlprefix }}%
\providecommand \urlprefix  [0]{URL }%
\providecommand \Eprint [0]{\href }%
\providecommand \doibase [0]{http://dx.doi.org/}%
\providecommand \selectlanguage [0]{\@gobble}%
\providecommand \bibinfo  [0]{\@secondoftwo}%
\providecommand \bibfield  [0]{\@secondoftwo}%
\providecommand \translation [1]{[#1]}%
\providecommand \BibitemOpen [0]{}%
\providecommand \bibitemStop [0]{}%
\providecommand \bibitemNoStop [0]{.\EOS\space}%
\providecommand \EOS [0]{\spacefactor3000\relax}%
\providecommand \BibitemShut  [1]{\csname bibitem#1\endcsname}%
\let\auto@bib@innerbib\@empty
\bibitem [{\citenamefont {Shang}\ \emph {et~al.}(2020)\citenamefont {Shang},
  \citenamefont {Zhang}, \citenamefont {Guo},\ and\ \citenamefont
  {Lü}}]{shang_heat_2020}%
  \BibitemOpen
  \bibfield  {author} {\bibinfo {author} {\bibfnamefont {M.-Y.}\ \bibnamefont
  {Shang}}, \bibinfo {author} {\bibfnamefont {C.}~\bibnamefont {Zhang}},
  \bibinfo {author} {\bibfnamefont {Z.}~\bibnamefont {Guo}}, \ and\ \bibinfo
  {author} {\bibfnamefont {J.-T.}\ \bibnamefont {Lü}},\ }\href {\doibase
  10.1038/s41598-020-65221-8} {\bibfield  {journal} {\bibinfo  {journal} {Sci.
  Rep.}\ }\textbf {\bibinfo {volume} {10}},\ \bibinfo {pages} {8272} (\bibinfo
  {year} {2020})}\BibitemShut {NoStop}%
\bibitem [{\citenamefont {Li}\ and\ \citenamefont
  {Lee}(2019)}]{PhysRevB.99.085202}%
  \BibitemOpen
  \bibfield  {author} {\bibinfo {author} {\bibfnamefont {X.}~\bibnamefont
  {Li}}\ and\ \bibinfo {author} {\bibfnamefont {S.}~\bibnamefont {Lee}},\
  }\href {\doibase 10.1103/PhysRevB.99.085202} {\bibfield  {journal} {\bibinfo
  {journal} {Phys. Rev. B}\ }\textbf {\bibinfo {volume} {99}},\ \bibinfo
  {pages} {085202} (\bibinfo {year} {2019})}\BibitemShut {NoStop}%
\bibitem [{\citenamefont {Guo}\ and\ \citenamefont
  {Wang}(2017)}]{wangmr17callaway}%
  \BibitemOpen
  \bibfield  {author} {\bibinfo {author} {\bibfnamefont {Y.}~\bibnamefont
  {Guo}}\ and\ \bibinfo {author} {\bibfnamefont {M.}~\bibnamefont {Wang}},\
  }\href {\doibase 10.1103/PhysRevB.96.134312} {\bibfield  {journal} {\bibinfo
  {journal} {Phys. Rev. B}\ }\textbf {\bibinfo {volume} {96}},\ \bibinfo
  {pages} {134312} (\bibinfo {year} {2017})}\BibitemShut {NoStop}%
\bibitem [{\citenamefont {Luo}\ \emph {et~al.}(2019)\citenamefont {Luo},
  \citenamefont {Guo}, \citenamefont {Wang},\ and\ \citenamefont
  {Yi}}]{luo2019}%
  \BibitemOpen
  \bibfield  {author} {\bibinfo {author} {\bibfnamefont {X.-P.}\ \bibnamefont
  {Luo}}, \bibinfo {author} {\bibfnamefont {Y.-Y.}\ \bibnamefont {Guo}},
  \bibinfo {author} {\bibfnamefont {M.-R.}\ \bibnamefont {Wang}}, \ and\
  \bibinfo {author} {\bibfnamefont {H.-L.}\ \bibnamefont {Yi}},\ }\href
  {\doibase 10.1103/PhysRevB.100.155401} {\bibfield  {journal} {\bibinfo
  {journal} {Phys. Rev. B}\ }\textbf {\bibinfo {volume} {100}},\ \bibinfo
  {pages} {155401} (\bibinfo {year} {2019})}\BibitemShut {NoStop}%
\bibitem [{\citenamefont {Cepellotti}\ \emph {et~al.}(2015)\citenamefont
  {Cepellotti}, \citenamefont {Fugallo}, \citenamefont {Paulatto},
  \citenamefont {Lazzeri}, \citenamefont {Mauri},\ and\ \citenamefont
  {Marzari}}]{cepellotti_phonon_2015}%
  \BibitemOpen
  \bibfield  {author} {\bibinfo {author} {\bibfnamefont {A.}~\bibnamefont
  {Cepellotti}}, \bibinfo {author} {\bibfnamefont {G.}~\bibnamefont {Fugallo}},
  \bibinfo {author} {\bibfnamefont {L.}~\bibnamefont {Paulatto}}, \bibinfo
  {author} {\bibfnamefont {M.}~\bibnamefont {Lazzeri}}, \bibinfo {author}
  {\bibfnamefont {F.}~\bibnamefont {Mauri}}, \ and\ \bibinfo {author}
  {\bibfnamefont {N.}~\bibnamefont {Marzari}},\ }\href {\doibase
  10.1038/ncomms7400} {\bibfield  {journal} {\bibinfo  {journal} {Nat.
  Commun.}\ }\textbf {\bibinfo {volume} {6}},\ \bibinfo {pages} {6400}
  (\bibinfo {year} {2015})}\BibitemShut {NoStop}%
\bibitem [{\citenamefont {Lee}\ \emph {et~al.}(2015)\citenamefont {Lee},
  \citenamefont {Broido}, \citenamefont {Esfarjani},\ and\ \citenamefont
  {Chen}}]{lee_hydrodynamic_2015}%
  \BibitemOpen
  \bibfield  {author} {\bibinfo {author} {\bibfnamefont {S.}~\bibnamefont
  {Lee}}, \bibinfo {author} {\bibfnamefont {D.}~\bibnamefont {Broido}},
  \bibinfo {author} {\bibfnamefont {K.}~\bibnamefont {Esfarjani}}, \ and\
  \bibinfo {author} {\bibfnamefont {G.}~\bibnamefont {Chen}},\ }\href {\doibase
  10.1038/ncomms7290} {\bibfield  {journal} {\bibinfo  {journal} {Nat.
  Commun.}\ }\textbf {\bibinfo {volume} {6}},\ \bibinfo {pages} {6290}
  (\bibinfo {year} {2015})}\BibitemShut {NoStop}%
\bibitem [{\citenamefont {Guyer}\ and\ \citenamefont
  {Krumhansl}(1966{\natexlab{a}})}]{PhysRev_GK}%
  \BibitemOpen
  \bibfield  {author} {\bibinfo {author} {\bibfnamefont {R.~A.}\ \bibnamefont
  {Guyer}}\ and\ \bibinfo {author} {\bibfnamefont {J.~A.}\ \bibnamefont
  {Krumhansl}},\ }\href {\doibase 10.1103/PhysRev.148.778} {\bibfield
  {journal} {\bibinfo  {journal} {Phys. Rev.}\ }\textbf {\bibinfo {volume}
  {148}},\ \bibinfo {pages} {778} (\bibinfo {year}
  {1966}{\natexlab{a}})}\BibitemShut {NoStop}%
\bibitem [{\citenamefont {Murthy}\ \emph {et~al.}(2005)\citenamefont {Murthy},
  \citenamefont {Narumanchi}, \citenamefont {Pascual-Gutierrez}, \citenamefont
  {Wang}, \citenamefont {Ni},\ and\ \citenamefont {Mathur}}]{MurthyJY05Review}%
  \BibitemOpen
  \bibfield  {author} {\bibinfo {author} {\bibfnamefont {J.~Y.}\ \bibnamefont
  {Murthy}}, \bibinfo {author} {\bibfnamefont {S.~V.~J.}\ \bibnamefont
  {Narumanchi}}, \bibinfo {author} {\bibfnamefont {J.~A.}\ \bibnamefont
  {Pascual-Gutierrez}}, \bibinfo {author} {\bibfnamefont {T.}~\bibnamefont
  {Wang}}, \bibinfo {author} {\bibfnamefont {C.}~\bibnamefont {Ni}}, \ and\
  \bibinfo {author} {\bibfnamefont {S.~R.}\ \bibnamefont {Mathur}},\ }\href
  {\doibase 10.1615/IntJMultCompEng.v3.i1.20} {\bibfield  {journal} {\bibinfo
  {journal} {Int. J. Multiscale Computat. Eng.}\ }\textbf {\bibinfo {volume}
  {3}},\ \bibinfo {pages} {5} (\bibinfo {year} {2005})}\BibitemShut {NoStop}%
\bibitem [{\citenamefont {Chen}(2005)}]{ChenG05Oxford}%
  \BibitemOpen
  \bibfield  {author} {\bibinfo {author} {\bibfnamefont {G.}~\bibnamefont
  {Chen}},\ }\href
  {https://global.oup.com/ushe/product/nanoscale-energy-transport-and-conversion-9780195159424?cc=cn&lang=en&}
  {\emph {\bibinfo {title} {Nanoscale energy transport and conversion: {A}
  parallel treatment of electrons, molecules, phonons, and photons}}}\
  (\bibinfo  {publisher} {Oxford University Press},\ \bibinfo {year}
  {2005})\BibitemShut {NoStop}%
\bibitem [{\citenamefont {Lee}\ and\ \citenamefont {Lindsay}(2017)}]{lee2017}%
  \BibitemOpen
  \bibfield  {author} {\bibinfo {author} {\bibfnamefont {S.}~\bibnamefont
  {Lee}}\ and\ \bibinfo {author} {\bibfnamefont {L.}~\bibnamefont {Lindsay}},\
  }\href {\doibase 10.1103/PhysRevB.95.184304} {\bibfield  {journal} {\bibinfo
  {journal} {Phys. Rev. B}\ }\textbf {\bibinfo {volume} {95}},\ \bibinfo
  {pages} {184304} (\bibinfo {year} {2017})}\BibitemShut {NoStop}%
\bibitem [{\citenamefont {Yang}\ \emph {et~al.}(2019)\citenamefont {Yang},
  \citenamefont {Yue},\ and\ \citenamefont {Liao}}]{Nanalytical}%
  \BibitemOpen
  \bibfield  {author} {\bibinfo {author} {\bibfnamefont {R.}~\bibnamefont
  {Yang}}, \bibinfo {author} {\bibfnamefont {S.}~\bibnamefont {Yue}}, \ and\
  \bibinfo {author} {\bibfnamefont {B.}~\bibnamefont {Liao}},\ }\href {\doibase
  10.1080/15567265.2018.1551449} {\bibfield  {journal} {\bibinfo  {journal}
  {Nanosc. Microsc. Therm.}\ }\textbf {\bibinfo {volume} {23}},\ \bibinfo
  {pages} {25} (\bibinfo {year} {2019})}\BibitemShut {NoStop}%
\bibitem [{\citenamefont {Li}\ and\ \citenamefont {Cao}(2019)}]{li2019}%
  \BibitemOpen
  \bibfield  {author} {\bibinfo {author} {\bibfnamefont {H.-L.}\ \bibnamefont
  {Li}}\ and\ \bibinfo {author} {\bibfnamefont {B.-Y.}\ \bibnamefont {Cao}},\
  }\href {\doibase 10.1080/15567265.2018.1520763} {\bibfield  {journal}
  {\bibinfo  {journal} {Nanosc. Microsc. Therm.}\ }\textbf {\bibinfo {volume}
  {23}},\ \bibinfo {pages} {10} (\bibinfo {year} {2019})}\BibitemShut {NoStop}%
\bibitem [{\citenamefont {Guyer}\ and\ \citenamefont
  {Krumhansl}(1966{\natexlab{b}})}]{PhysRev.148.766}%
  \BibitemOpen
  \bibfield  {author} {\bibinfo {author} {\bibfnamefont {R.~A.}\ \bibnamefont
  {Guyer}}\ and\ \bibinfo {author} {\bibfnamefont {J.~A.}\ \bibnamefont
  {Krumhansl}},\ }\href {\doibase 10.1103/PhysRev.148.766} {\bibfield
  {journal} {\bibinfo  {journal} {Phys. Rev.}\ }\textbf {\bibinfo {volume}
  {148}},\ \bibinfo {pages} {766} (\bibinfo {year}
  {1966}{\natexlab{b}})}\BibitemShut {NoStop}%
\bibitem [{\citenamefont {Zhang}\ \emph {et~al.}(2019)\citenamefont {Zhang},
  \citenamefont {Guo},\ and\ \citenamefont {Chen}}]{ZHANG20191366}%
  \BibitemOpen
  \bibfield  {author} {\bibinfo {author} {\bibfnamefont {C.}~\bibnamefont
  {Zhang}}, \bibinfo {author} {\bibfnamefont {Z.}~\bibnamefont {Guo}}, \ and\
  \bibinfo {author} {\bibfnamefont {S.}~\bibnamefont {Chen}},\ }\href {\doibase
  10.1016/j.ijheatmasstransfer.2018.10.141} {\bibfield  {journal} {\bibinfo
  {journal} {Int. J. Heat Mass Transfer}\ }\textbf {\bibinfo {volume} {130}},\
  \bibinfo {pages} {1366} (\bibinfo {year} {2019})}\BibitemShut {NoStop}%
\bibitem [{\citenamefont {Van~Leer}(1977)}]{vanleer1977}%
  \BibitemOpen
  \bibfield  {author} {\bibinfo {author} {\bibfnamefont {B.}~\bibnamefont
  {Van~Leer}},\ }\href {\doibase 10.1016/0021-9991(77)90095-X} {\bibfield
  {journal} {\bibinfo  {journal} {J. Comput. Phys.}\ }\textbf {\bibinfo
  {volume} {23}},\ \bibinfo {pages} {276} (\bibinfo {year} {1977})}\BibitemShut
  {NoStop}%
\bibitem [{\citenamefont {Hale}\ and\ \citenamefont
  {Townsend}(2013)}]{NicholasH13GaussL}%
  \BibitemOpen
  \bibfield  {author} {\bibinfo {author} {\bibfnamefont {N.}~\bibnamefont
  {Hale}}\ and\ \bibinfo {author} {\bibfnamefont {A.}~\bibnamefont
  {Townsend}},\ }\href {\doibase 10.1137/120889873} {\bibfield  {journal}
  {\bibinfo  {journal} {Siam J. Sci. Comput}\ }\textbf {\bibinfo {volume}
  {35}},\ \bibinfo {pages} {A652} (\bibinfo {year} {2013})}\BibitemShut
  {NoStop}%
\bibitem [{\citenamefont {Olfe}(1968)}]{olfe1968}%
  \BibitemOpen
  \bibfield  {author} {\bibinfo {author} {\bibfnamefont {D.~B.}\ \bibnamefont
  {Olfe}},\ }\href {\doibase 10.1016/0022-4073(68)90094-0} {\bibfield
  {journal} {\bibinfo  {journal} {J. Quant. Spectrosc. Ra.}\ }\textbf {\bibinfo
  {volume} {8}},\ \bibinfo {pages} {899} (\bibinfo {year} {1968})}\BibitemShut
  {NoStop}%
\bibitem [{\citenamefont {Majumdar}(1993)}]{MajumdarA93Film}%
  \BibitemOpen
  \bibfield  {author} {\bibinfo {author} {\bibfnamefont {A.}~\bibnamefont
  {Majumdar}},\ }\href {\doibase 10.1115/1.2910673} {\bibfield  {journal}
  {\bibinfo  {journal} {J. Heat Transfer}\ }\textbf {\bibinfo {volume} {115}},\
  \bibinfo {pages} {7} (\bibinfo {year} {1993})}\BibitemShut {NoStop}%
\bibitem [{\citenamefont {Chen}(1996)}]{chen1996}%
  \BibitemOpen
  \bibfield  {author} {\bibinfo {author} {\bibfnamefont {G.}~\bibnamefont
  {Chen}},\ }\href {\doibase 10.1115/1.2822665} {\bibfield  {journal} {\bibinfo
   {journal} {J. Heat Transfer}\ }\textbf {\bibinfo {volume} {118}},\ \bibinfo
  {pages} {539} (\bibinfo {year} {1996})}\BibitemShut {NoStop}%
\bibitem [{\citenamefont {Pop}\ \emph {et~al.}(2012)\citenamefont {Pop},
  \citenamefont {Varshney},\ and\ \citenamefont {Roy}}]{pop_varshney_roy_2012}%
  \BibitemOpen
  \bibfield  {author} {\bibinfo {author} {\bibfnamefont {E.}~\bibnamefont
  {Pop}}, \bibinfo {author} {\bibfnamefont {V.}~\bibnamefont {Varshney}}, \
  and\ \bibinfo {author} {\bibfnamefont {A.~K.}\ \bibnamefont {Roy}},\ }\href
  {\doibase 10.1557/mrs.2012.203} {\bibfield  {journal} {\bibinfo  {journal}
  {MRS Bull.}\ }\textbf {\bibinfo {volume} {37}},\ \bibinfo {pages} {1273}
  (\bibinfo {year} {2012})}\BibitemShut {NoStop}%
\bibitem [{\citenamefont {Yang}\ \emph {et~al.}(2015)\citenamefont {Yang},
  \citenamefont {Hu}, \citenamefont {Ma}, \citenamefont {Lu},\ and\
  \citenamefont {Li}}]{yang_nanoscale_2015}%
  \BibitemOpen
  \bibfield  {author} {\bibinfo {author} {\bibfnamefont {N.}~\bibnamefont
  {Yang}}, \bibinfo {author} {\bibfnamefont {S.}~\bibnamefont {Hu}}, \bibinfo
  {author} {\bibfnamefont {D.}~\bibnamefont {Ma}}, \bibinfo {author}
  {\bibfnamefont {T.}~\bibnamefont {Lu}}, \ and\ \bibinfo {author}
  {\bibfnamefont {B.}~\bibnamefont {Li}},\ }\href {\doibase 10.1038/srep14878}
  {\bibfield  {journal} {\bibinfo  {journal} {Sci. Rep.}\ }\textbf {\bibinfo
  {volume} {5}},\ \bibinfo {pages} {14878} (\bibinfo {year}
  {2015})}\BibitemShut {NoStop}%
\bibitem [{\citenamefont {Ma}\ \emph {et~al.}(2017)\citenamefont {Ma},
  \citenamefont {Ding}, \citenamefont {Wang}, \citenamefont {Yang},\ and\
  \citenamefont {Zhang}}]{ma_unexpected_2017}%
  \BibitemOpen
  \bibfield  {author} {\bibinfo {author} {\bibfnamefont {D.}~\bibnamefont
  {Ma}}, \bibinfo {author} {\bibfnamefont {H.}~\bibnamefont {Ding}}, \bibinfo
  {author} {\bibfnamefont {X.}~\bibnamefont {Wang}}, \bibinfo {author}
  {\bibfnamefont {N.}~\bibnamefont {Yang}}, \ and\ \bibinfo {author}
  {\bibfnamefont {X.}~\bibnamefont {Zhang}},\ }\href {\doibase
  https://doi.org/10.1016/j.ijheatmasstransfer.2016.12.092} {\bibfield
  {journal} {\bibinfo  {journal} {Int. J. Heat Mass Transfer}\ }\textbf
  {\bibinfo {volume} {108}},\ \bibinfo {pages} {940 } (\bibinfo {year}
  {2017})}\BibitemShut {NoStop}%
\bibitem [{\citenamefont {Barenblatt}(1987)}]{barenblatt1987dimensional}%
  \BibitemOpen
  \bibfield  {author} {\bibinfo {author} {\bibfnamefont {G.~I.}\ \bibnamefont
  {Barenblatt}},\ }\href@noop {} {\emph {\bibinfo {title} {Dimensional
  analysis}}}\ (\bibinfo  {publisher} {CRC Press},\ \bibinfo {year}
  {1987})\BibitemShut {NoStop}%
\end{thebibliography}%

\end{document}


\title{Supplemental Material: Violation of Fourier's law in homogeneous systems}
\author{Chuang Zhang}
\affiliation{State Key Laboratory of Coal Combustion, School of Energy and Power Engineering, Huazhong University of Science and Technology,Wuhan, 430074, China}
\author{Dengke Ma}
\affiliation{NNU-SULI Thermal Energy Research Center (NSTER) and Center for Quantum Transport and Thermal Energy Science (CQTES), School of Physics and Technology, Nanjing Normal University, Nanjing, 210023, China}
\author{Manyu Shang}
\affiliation{School of Physics and Wuhan National High Magnetic Field Center, Huazhong University of Science and Technology, Wuhan 430074, P. R. China}
\author{Xiao Wan}
\affiliation{State Key Laboratory of Coal Combustion, School of Energy and Power Engineering, Huazhong University of Science and Technology,Wuhan, 430074, China}
\author{Jing-Tao L\"u}
\affiliation{School of Physics and Wuhan National High Magnetic Field Center, Huazhong University of Science and Technology, Wuhan 430074, P. R. China}
\author{Zhaoli Guo}
\email{Corresponding author: zlguo@hust.edu.cn}
\affiliation{State Key Laboratory of Coal Combustion, School of Energy and Power Engineering, Huazhong University of Science and Technology,Wuhan, 430074, China}
\author{Baowen Li}
\email{Corresponding author: Baowen.Li@Colorado.Edu}
\affiliation{Paul M Rady Department of Mechanical Engineering, Department of Physics, University of Colorado, Boulder, Colorado 80309, USA}
\author{Nuo Yang}
\email{Corresponding author: nuo@hust.edu.cn}
\affiliation{State Key Laboratory of Coal Combustion, School of Energy and Power Engineering, Huazhong University of Science and Technology,Wuhan, 430074, China}

\date{\today}
\maketitle

\section{Phonon Boltzmann transport equation}
\label{sec:modeleq}

{\color{black}{In this section, the stationary Phonon Boltzmann transport equation (BTE) under the Callaway's dual relaxation model~\cite{shang_heat_2020,PhysRevB.99.085202,wangmr17callaway,luo2019} is introduced in detail.}}

This model equation is~\cite{shang_heat_2020,PhysRevB.99.085202,cepellotti_phonon_2015,lee_hydrodynamic_2015,wangmr17callaway,luo2019,PhysRev_GK},
\begin{equation}
 \bm{v} \cdot \nabla f= \frac{f^{eq}_{R}-f}{\tau_{R}} + \frac{f^{eq}_{N}-f}{\tau_{N}},
\label{eq:BTE}
\end{equation}
where $f=f(\bm{x},\bm{K},\omega, p )$ is the phonon distribution function, $\bm{x}$ is the physical position, $\bm{K}$ is the wave vector and assumed to be isotropic, i.e., $\bm{K}=|\bm{K}| \bm{s}$, $\bm{s}$ is the unit directional vector, $\omega$ is the angular frequency, $\bm{v}=\nabla_{\bm{K}} \omega$ is the group velocity, $p$ is the phonon polarization.
The left side of Eq.~\eqref{eq:BTE} represents the phonon advection and the right side of Eq.~\eqref{eq:BTE} is the phonon scattering term~\cite{MurthyJY05Review}, which is composed of two parts: the first part is the momentum destroying resistive (R) scattering and the second part is the momentum conservation normal (N) scattering~\cite{cepellotti_phonon_2015,lee_hydrodynamic_2015,wangmr17callaway}.
$\tau_{R}$ is the effective relaxation time of the R scattering, which is a combination of all momentum destroying phonon scattering except the boundary scattering based on the Mathiessen's rule~\cite{ChenG05Oxford}.
$f^{eq}_{R}$ is the associated equilibrium state of the R scattering and satisfies Bose-Einstein distribution~\cite{ChenG05Oxford}, i.e.,
\begin{align}
f^{eq}_{R}(T)  = \frac{1}{ \exp \left( \frac{\hbar \omega }{k_B T} \right) -1 },  \label{eq:feqresistive}
\end{align}
where $k_B$ and $\hbar$ are the Boltzmann constant and Planck constant reduced by $2\pi$, $T$ is the temperature.
Different from R scattering, the N scattering satisfies the momentum conservation.
Its displaced equilibrium distribution function and the effective relaxation time are $f^{eq}_{N}$ and $\tau_{N}$, respectively, where
\begin{align}
f^{eq}_{N}(T,\bm{u})= \frac{1}{ \exp \left( \frac{\hbar \omega - \hbar \bm{K} \cdot \bm{u} }{k_B T} \right) -1 }, \label{eq:feqnormal}
\end{align}
where $\bm{u}$ is the drift velocity.

Usually, the phonon BTE (Eq.~\eqref{eq:BTE}) can be rewritten into a deviational energy form as below
\begin{align}
\bm{v} \cdot \nabla e= \frac{e^{eq}_{R}-e}{\tau_{R}} + \frac{e^{eq}_{N}-e}{\tau_{N}},
\label{eq:energyBTE}
\end{align}
where the associated deviational distribution functions of energy density are~\cite{luo2019,wangmr17callaway}
\begin{align}
e &=\frac{ \hbar \omega  D (f-f_R^{eq} (T_0) ) }{A}, \\
e_R^{eq} &=\frac{ \hbar \omega  D (f_R^{eq}-f_R^{eq} (T_0) ) }{A}, \label{eq:eeqresistive} \\
e_N^{eq} &=\frac{ \hbar \omega  D (f_N^{eq}-f_R^{eq} (T_0) ) }{A}, \label{eq:eeqnormal}
\end{align}
where $T_0$ is the reference temperature, $D(\omega,p)$ is the phonon density of state~\cite{ChenG05Oxford,wangmr17callaway}.
In 2D systems, $A=2\pi$, $D=|\bm{K}|/\left( 2 \pi |\bm{v}| \right)$, in 3D systems, $A=4\pi$, $D=|\bm{K}|^2/\left( 2 {\pi}^2 |\bm{v}| \right)$.

In this study, assuming a small temperature difference and a small drift velocity, i.e., $\Delta T/T_0 \ll 1$, $\bm{K} \cdot \bm{u} \ll \omega$, so that the equilibrium distribution function can be linearized, i.e.,
\begin{align}
e^{eq}_{R}(T) &\approx C \left( T-T_0 \right) / A, \label{eq:feqR} \\
e^{eq}_{N}(T,\bm{u}) &\approx  C \left( T-T_0 \right) / A +CT \frac{\bm{K} \cdot \bm{u} }{ A \omega},\label{eq:feqN}
\end{align}
where $C=C(\omega,p,T_0)$ is the mode specific heat at $T_0$, i.e.,
\begin{equation}
C(\omega,p,T_0)=A  \left.\ \frac{  \partial{ e^{eq}_{R}}}{ \partial{T}} \right|_{T=T_0} .
\label{eq:specificheat}
\end{equation}
The local deviational energy $E$, local temperature $T$ and local heat flux $\bm{q}$ can be updated by taking the moment of the distribution function, i.e.,
\begin{align}
E &= \sum_{p} \int \int e d\Omega d\omega,  \label{eq:energy} \\
T  &=T_0+ \frac{ \sum_{p} \int \int e d\Omega d\omega } { \sum_{p} \int C d\omega  },   \label{eq:T}  \\
\bm{q} &=  \sum_{p} \int \int \bm{v} e d\Omega d\omega ,
\label{eq:heatflux}
\end{align}
where $d\Omega$ and $d\omega$ are the integral over the whole 2D (or 3D) solid angle space and frequency space.

The N and R scattering satisfy the energy conservation, i.e.,
\begin{align}
0  &=  \sum_{p} \int \int    \frac{e^{eq}_{N} (T_N)-e}{\tau_{N}}   d\Omega d\omega , \label{eq:Nconsertvation}  \\
0  &=  \sum_{p} \int \int  \frac{e^{eq}_{R}(T_R) -e}{\tau_{R}}   d\Omega d\omega , \label{eq:Rconsertvation}
\end{align}
where $T_R$ and $T_N$ are local pseudotemperatures, which are introduced to ensure the conservation principle of the phonon scattering.
In addition, the N scattering satisfies the momentum conservation, i.e.,
\begin{align}
0 =  \sum_{p} \int \int  \frac{\bm{K} }{\omega } \frac{e^{eq}_{N} (T_N,\bm{u})-e}{\tau_{N}}   d\Omega d\omega . \label{eq:NMconsertvation}
\end{align}
Based on Eqs.~\eqref{eq:Nconsertvation},~\eqref{eq:Rconsertvation} and~\eqref{eq:NMconsertvation}, the macroscopic variables $T_R$, $T_N$ and $\bm{u}$ can be obtained,
\begin{align}
T_{R} & =T_{0}+ \left( \sum_{p}\int  \frac{\int e d\Omega }{\tau_R } d{\omega}   \right) \times \left(  \sum_{p}\int \frac{ C}{\tau_R }  d{\omega} \right)^{-1} , \\
T_{N} & =T_{0}+ \left( \sum_{p}\int  \frac{\int e d\Omega }{\tau_N } d{\omega}   \right) \times \left(  \sum_{p}\int \frac{ C}{\tau_N }  d{\omega} \right)^{-1} , \\
\bm{u}  &=  \frac{2}{T_{N}  \sum_{p} \int \frac{ |\bm{K}|^2  }{ \omega^2  } \frac{ C }{\tau_N } d{\omega}  } \sum_{p} \int \int  \frac{ \bm{K} }{\omega} \frac{ e }{ \tau_N}   d\Omega d\omega.
\end{align}
Please note that the drift velocity is nonzero in the hydrodynamics regime~\cite{shang_heat_2020,lee2017,Nanalytical,PhysRevB.99.085202,wangmr17callaway,luo2019}.

In addition, boundary conditions also play an indispensable role on thermal conduction.
Here, the thermalizing boundary condition~\cite{li2019,luo2019,wangmr17callaway} is used to deal with the inner and outer heat baths with a fixed temperature $T_{\text{w}}$, i.e.,
\begin{align}
e(\bm{x}_b,\bm{s})=e_R^{eq}(T_{\text{w}}) \quad ( \text{or} ~ f(\bm{x}_b,\bm{s})=f_R^{eq}(T_{\text{w}})),    \quad \bm{s} \cdot \mathbf{n}_{b} >0,
\label{eq:BC}
\end{align}
where $\mathbf{n}_{b}$ is the unit normal vector of the boundary $\bm{x}_b$ pointing to the computational domain.
Equation~\eqref{eq:BC} indicates that the distribution functions of all phonons emitting from the inner (outer) heat bath are $e_R^{eq}$ with fixed heat bath temperature $T_h$ ($T_c$)~\cite{wangmr17callaway,luo2019}.

Based on Eqs.~\eqref{eq:T} and \eqref{eq:heatflux}, the local thermal conductivity along the radial direction in homogeneous 2D disk/3D sphere can be obtained, i.e.,
\begin{equation}
\kappa (r) = -\frac{q(r)} {dT/dr },   \quad l <  r  < L,
\end{equation}
where $q(r)$ is the local heat flux, namely, the heat energy flow along the radial direction per unit area in a unit time. $r$ is the distance from the center in homogeneous 2D disk/3D sphere.
Then the graded thermal conductivity in 2D disk/3D sphere in the ballistic, diffusive and phonon hydrodynamic~\cite{PhysRevB.99.085202,PhysRev_GK,PhysRev.148.766,Nanalytical,lee_hydrodynamic_2015,cepellotti_phonon_2015} regimes can be investigated numerically based on phonon BTE.

\section{Phonon dispersion and scattering in graphene}
\label{sec:graphenedispersion}

{\color{black}{The phonon dispersion and polarization of graphene are calculated using Vienna Ab initio Simulation Package (VASP) combined with phonopy.
The supercell size for both second and third order force
constant calculation is $5 \times 5 \times 1$. The numbers of mesh points for reciprocal space sampling are $151 \times 151 \times 1$. The phonon-phonon interaction and related properties can be calculated using phono3py.
The phonon dispersion of graphene along $\Gamma$-M direction is shown in~\cref{gammaMdispersion}.
The associated phonon N/R scattering rates along $\Gamma$-M direction in different temperatures are shown in~\cref{graphenescattering}, where the natural abundance of graphene is $1.1\%$ $^{13}$C.
\begin{figure}
     \centering
     \includegraphics[scale=0.3]{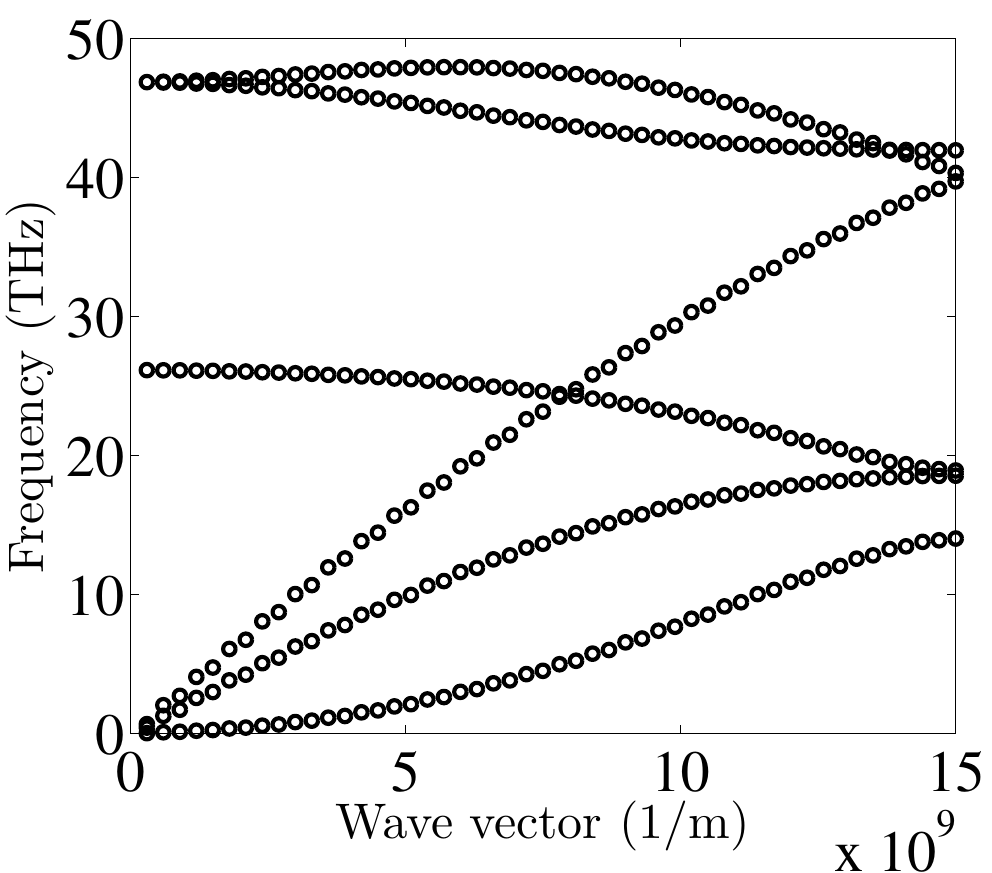}
     \caption{The phonon dispersion of graphene along $\Gamma$-M direction.}
     \label{gammaMdispersion}
\end{figure}
\begin{figure}
     \centering
     \subfloat[]{\includegraphics[scale=0.3]{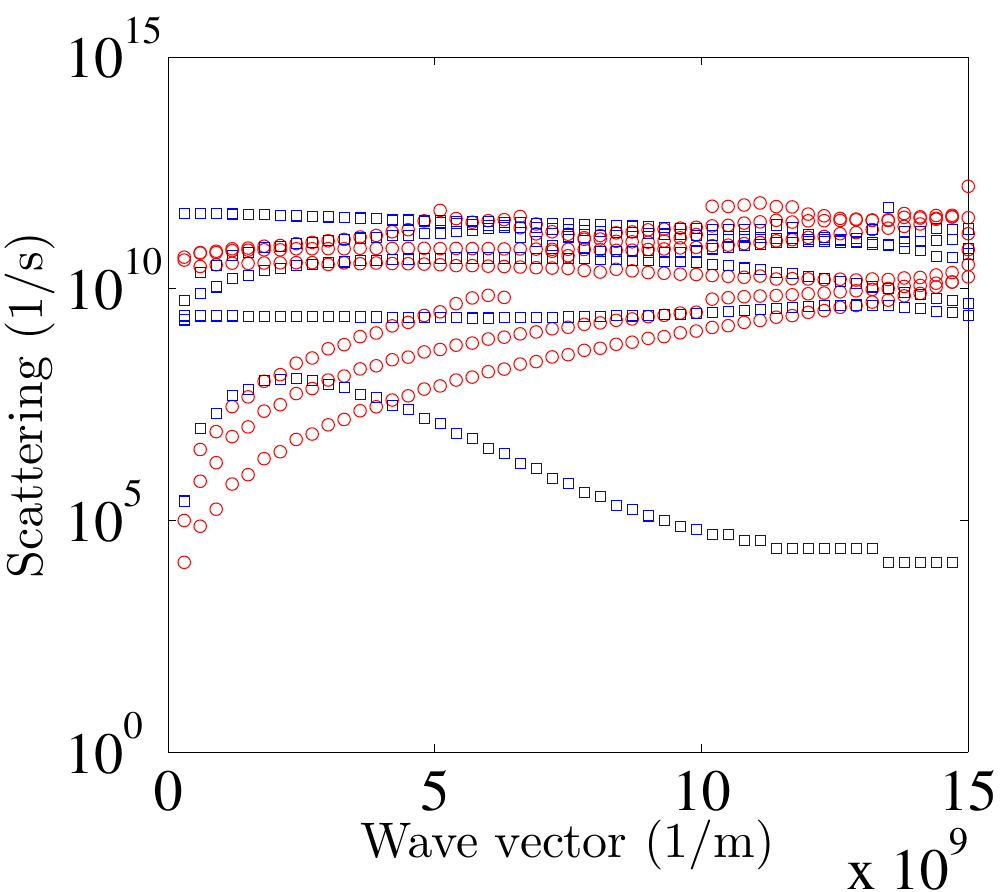}}~
     \subfloat[]{\includegraphics[scale=0.3]{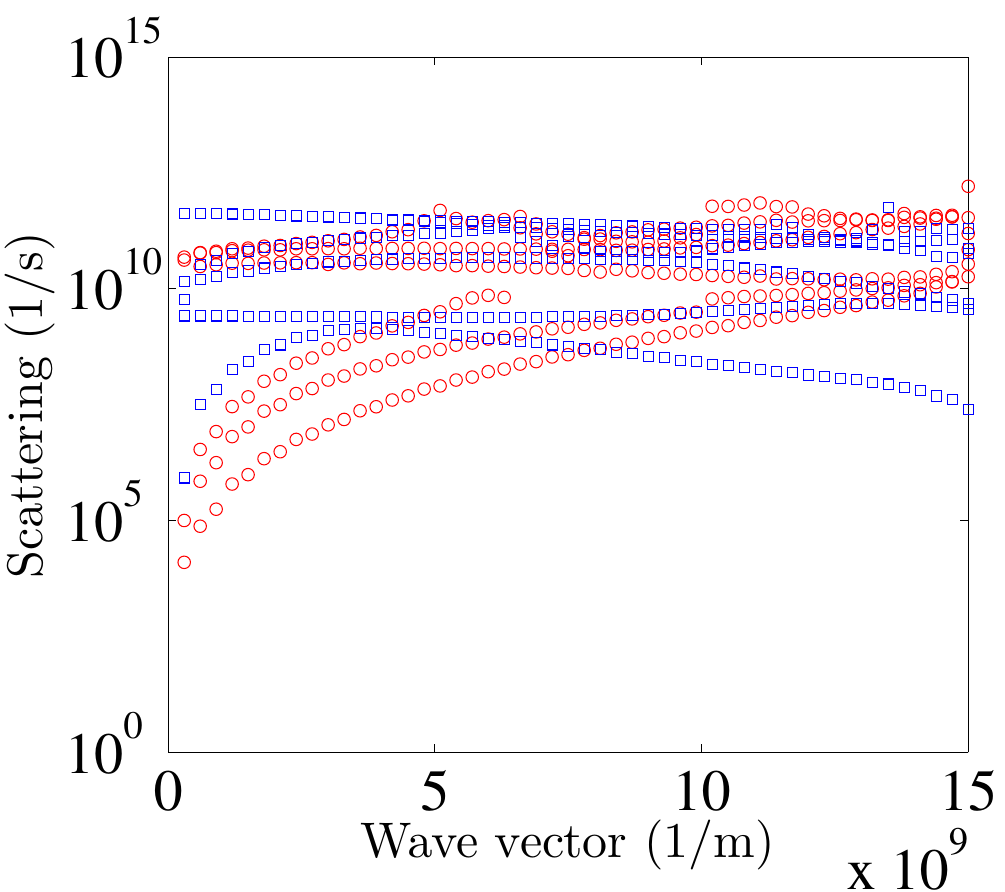}} \\
     \subfloat[]{\includegraphics[scale=0.3]{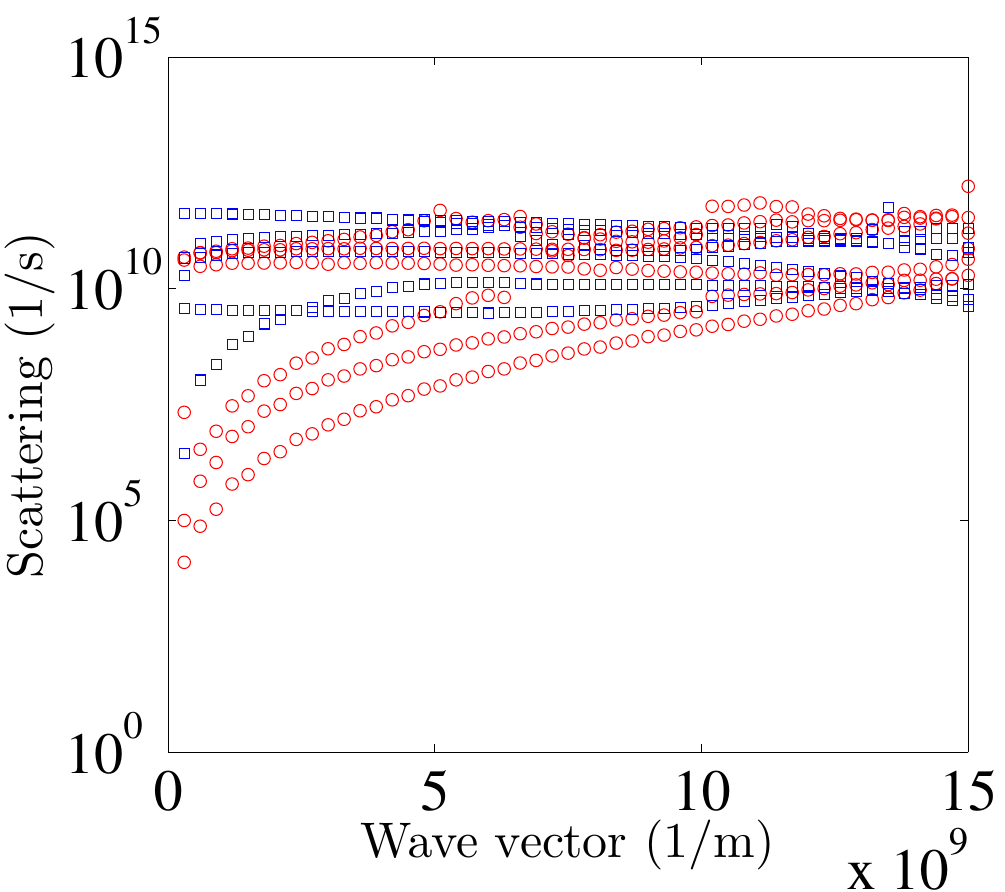}}~
     \subfloat[]{\includegraphics[scale=0.3]{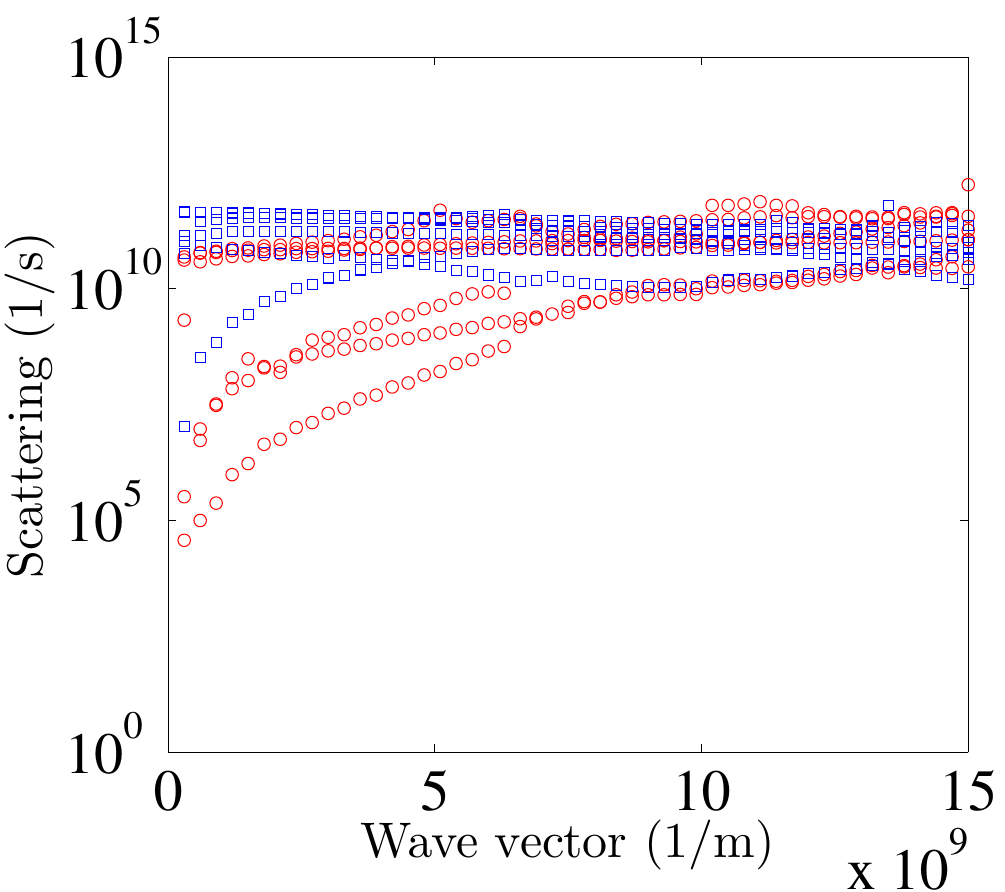}}\\
     \caption{Phonon N/R scattering rates of graphene (natural abundance) along $\Gamma$-M direction in different temperatures, where the red circles represent R scattering rates and the blue squares represent N scattering rates. (a) $10$ K, (b) $30$ K, (c) $100$ K, (d) $300$ K.}
     \label{graphenescattering}
\end{figure}
\begin{figure}
     \centering
     \includegraphics[scale=0.3]{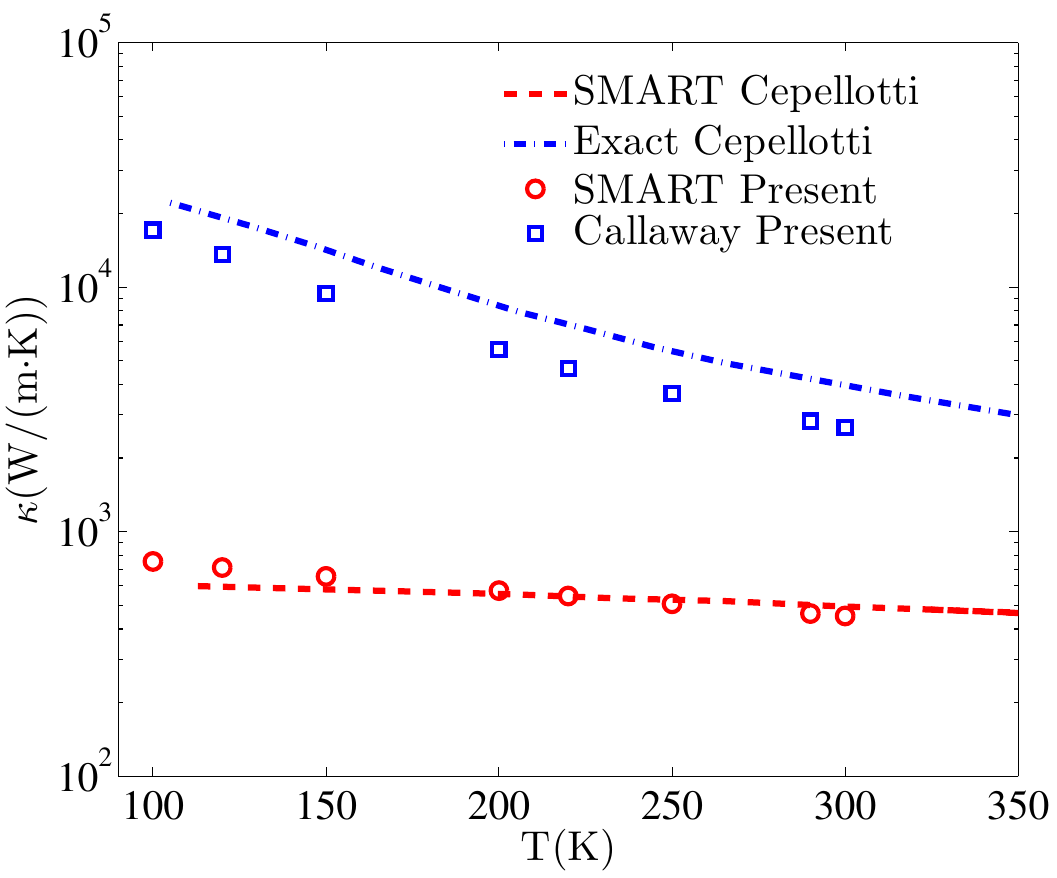}
     \caption{Thermal conductivity of graphene in different temperatures, where the system size is infinite. The symbols are the present results and the lines are the data obtained from Cepellotti's paper~\cite{cepellotti_phonon_2015}. Exact represents the variational solution of phonon BTE under ab initio full scattering kernel~\cite{cepellotti_phonon_2015}, SMART represents the solution of Phonon BTE under single mode relaxation time approximation, Callaway represents the solution of Phonon BTE under Callaway's model~\cite{wangmr17callaway,PhysRev_GK}.  }
     \label{kappaTgraphene}
\end{figure}

In our present simulations, the isotropic wave vector space is assumed to save computational time.
The phonon dispersion and thermal properties (\cref{gammaMdispersion},\cref{graphenescattering}) of graphene along $\Gamma$-M direction are used , which is accurate enough to predict the thermal conductivity of graphene in a wide temperature range.
The associated thermal conductivity of graphene in different temperatures are shown in~\cref{kappaTgraphene}, where the system size is infinite.
The thermal conductivity predicted by Callaway's model is $2674$ W/(m$\cdot$K) and $17095$ W/(m$\cdot$K) in $300$ K and $100$ K, respectively.
}}

\section{Numerical solutions}
\label{sec:discretization}

The implicit discrete ordinate method (DOM)~\cite{wangmr17callaway,ZHANG20191366} are used to solve the steady BTE, respectively.
To ensure the numerical accuracy, enough cells are used to discrete the high-dimensional phase space.

In 2D disk, the spatial space is discretized with $200 \times 200$ uniform cells and the van Leer limiter~\cite{vanleer1977} is used to ensure stability.
For the solid angle space, we set $\bm{s}=\left( \cos \theta, \sin \theta  \right)$, where $\theta \in [0, 2 \pi]$ is the polar angle.
Due to symmetry, the $\theta \in [0,\pi]$ is discretized with the $N_{\theta}$-point Gauss-Legendre quadrature~\cite{NicholasH13GaussL}.
The total number of the discretized directions is $2N_{\theta}$.
In our 2D simulations, we set $N_{\theta}=50$.
In graphene disk, the discretized cells and directions are the same as those in 2D disk.
The discretized phonon dispersion and polarization are shown in~\cref{gammaMdispersion} and the mid-point rule is used for the numerical integration of the frequency space.

In 3D sphere, the spatial space is discretized with $200 \times 200 \times 200$ uniform cells and the van Leer limiter is used, too.
For the solid angle space, $\bm{s}=\left( \cos \theta, \sin \theta \cos \varphi, \sin \theta  \sin \varphi  \right)$, where $\theta \in [0,\pi]$, $\varphi \in [0,2\pi]$ is the azimuthal angle.
The $\cos \theta \in [-1,1]$ is discretized with the $N_{\theta}$-point Gauss-Legendre quadrature~\cite{NicholasH13GaussL}, while the azimuthal angular space $\varphi \in [0,\pi]$ (due to symmetry) is discretized with the $\frac{N_{\varphi}}{2}$-point Gauss-Legendre quadrature.
In our 3D simulations, we set $N_{\theta} \times N_{\varphi} =24 \times 24$ or $36 \times 36$.

\section{Analytical solutions of the phonon BTE in the ballistic, diffusive and hydrodynamic limits}
\label{sec:analyticalsolutions}

The separate thermal effects of normal (N) scattering and resistive (R) scattering on graded thermal conductivity are investigated.
The analytical solutions in three limits at steady state are derived based on phonon BTE (Eq.\eqref{eq:energyBTE}) with rigorous mathematical derivations:
\begin{enumerate}
    \item ballistic limit (No N scattering, no R scattering),
    \item diffusive limit (No N scattering, very frequent R scattering),
    \item phonon hydrodynamic limit~\cite{PhysRevB.99.085202,PhysRev_GK,PhysRev.148.766,Nanalytical,lee_hydrodynamic_2015,cepellotti_phonon_2015} (No R scattering, very frequent N scattering).
\end{enumerate}

Without special statements, the Debye approximation and gray model~\cite{ChenG05Oxford,Nanalytical} are used, i.e., $|\bm{v}|= \omega/|\bm{K}|$.
The Knudsen number is introduced and defined as the ratio between the phonon mean free path $\lambda= |\bm{v}| \tau$ to the diameter of outer heat bath $(2L)$, i.e., $\text{Kn}=\lambda /2L$, where $\tau^{-1}=\tau^{-1}_N+\tau^{-1}_R$.
First, we derive the analytical solutions in 2D polar systems.
Then the mathematical derivations in 3D spherical systems are similar and straightforward.

\subsection{2D disk}

\begin{figure}[htb]
 \centering
 \includegraphics[scale=0.25,viewport=150 0 750 600,clip=true]{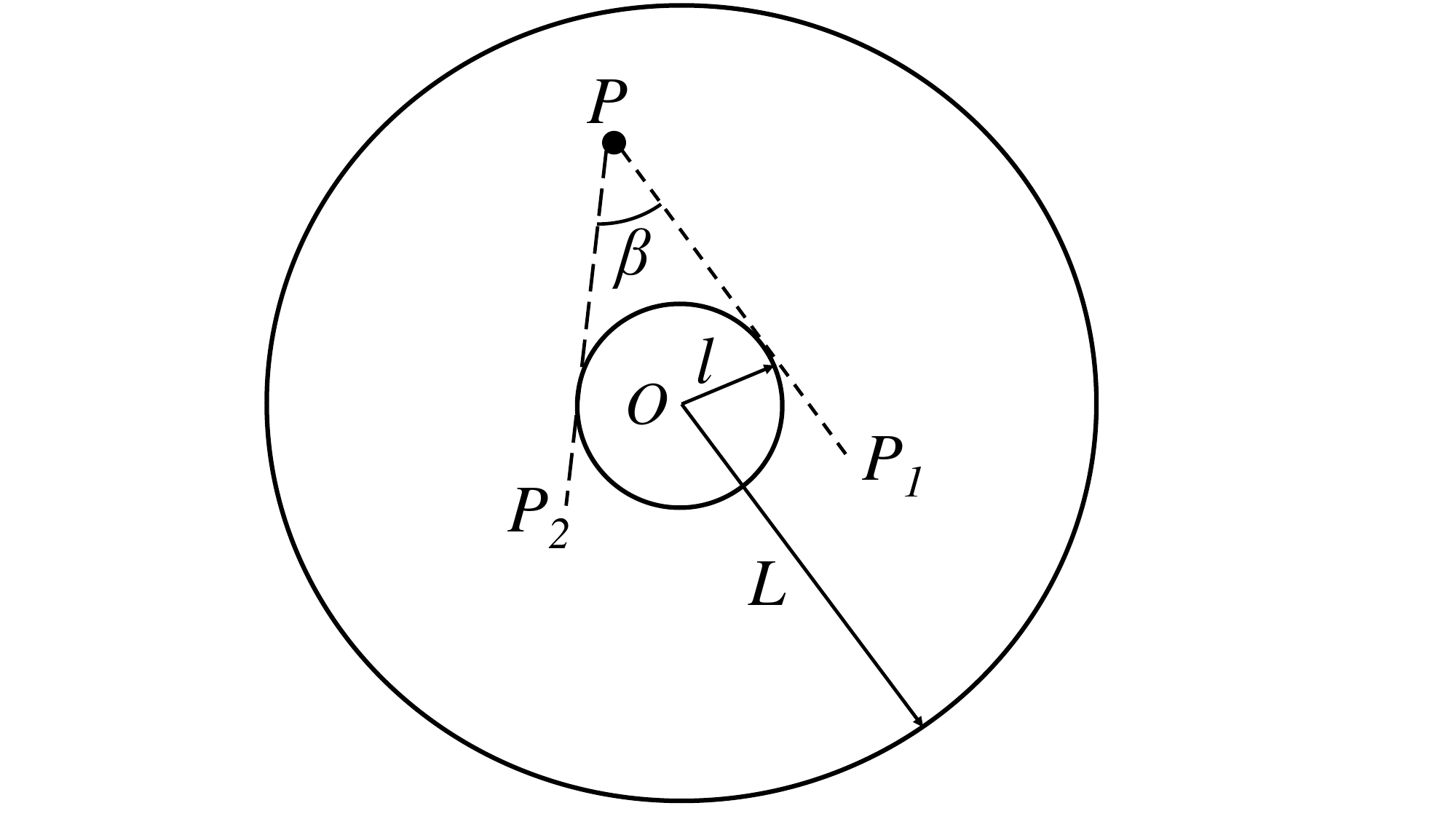}
 \caption{Heat conduction in homogeneous 2D disk in the ballistic limit at steady state. $O$ is the center of the geometry. The radii of the inner and outer heat baths (or boundaries) are $l,~L$, respectively. The temperatures of the inner and outer heat baths (boundaries) are set $T_h$ and $T_c$. Lines $PP_1$ and $PP_2$ are tangent to the inner heat bath. $\beta =2\arcsin (l/r)$, where $r$ is the distance from the disk center $(| OP |)$. }
 \label{ballisticpdf}
\end{figure}

\begin{figure}[htb]
 \centering
 \includegraphics[scale=0.25,viewport=100 50 1600 900,clip=true]{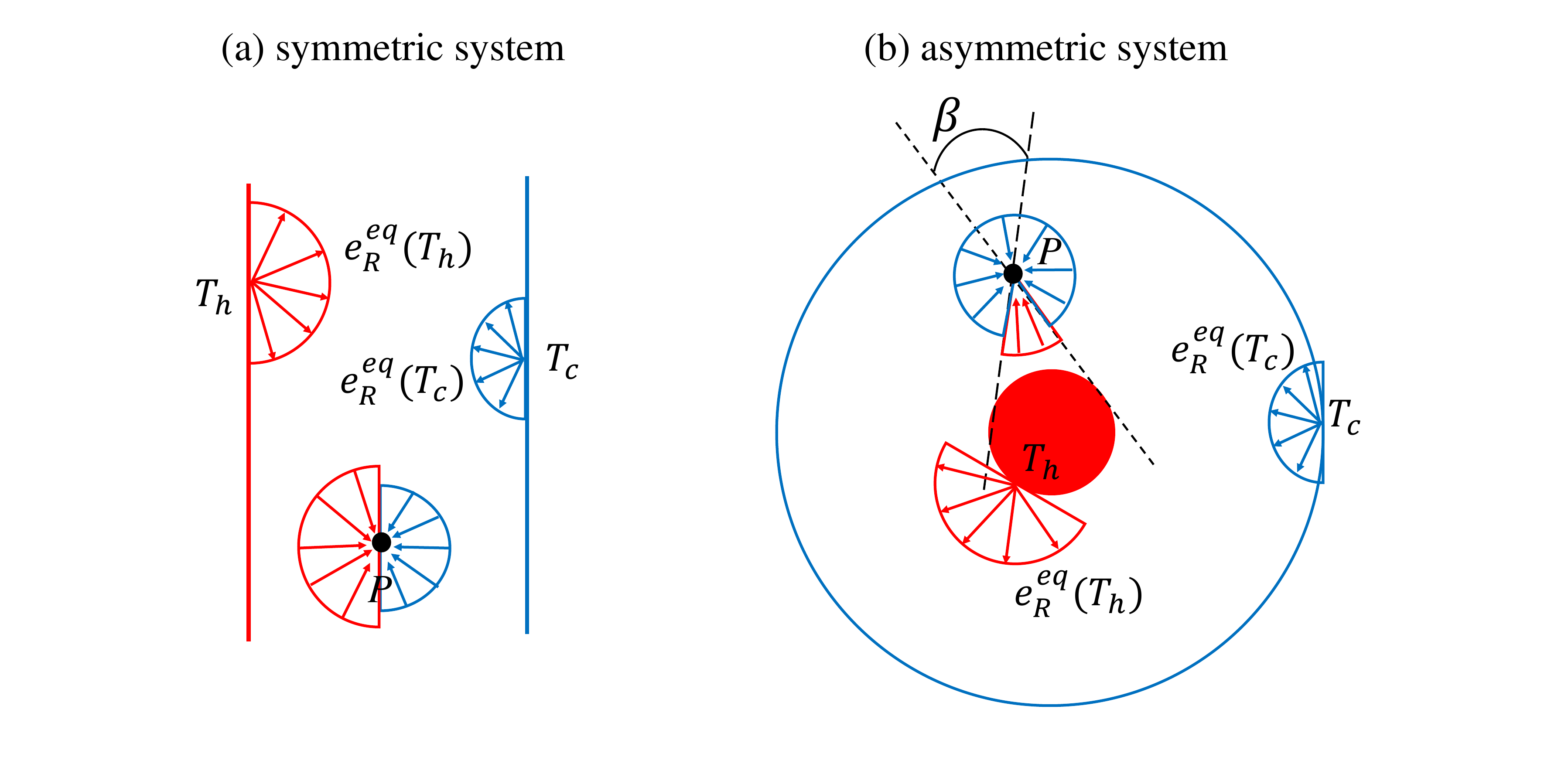}
 \caption{Ballistic phonon transport in (a) symmetric system~\cite{MajumdarA93Film} and (b) asymmetric system~\cite{chen1996,li2019} with two thermal baths ($T_h$, $T_c$). Based on isothermal boundary conditions (Eq.~\eqref{eq:BC}), the distribution functions of all phonons emitting from the hot and cold heat baths are $e^{eq}_{R} (T_h)$ and $e^{eq}_{R} (T_c)$, respectively~\cite{wangmr17callaway,luo2019}. For an arbitrary position $P$ in the interior domain, phonons going through this position come from the hot and cold thermal baths with different directions~\cite{chen1996,li2019}. (a) In the symmetric system, all phonons emitting from one heat bath will be totally received by the other. So that the temperature inside the system is a constant~\cite{MajumdarA93Film}. (b) In asymmetric system, for all phonons with arbitrary directions transporting through $P$, only a portion of them come from the hot heat bath with $e^{eq}_R(T_h)$, while the other come from the cold heat bath with $e^{eq}_R(T_c)$. So that the local temperature (Eq.~\eqref{eq:TP}) is not a constant, but related to $\beta =2\arcsin (l/r)$ (\cref{ballisticpdf}), where $r$ is the distance between the disk center and position $P$. }
 \label{ballisticasymmetric}
\end{figure}

As the system size is much smaller than the phonon mean free path, i.e., $\text{Kn} \rightarrow \infty$, the heat conduction is in the ballistic regime~\cite{li2019,olfe1968} and there is rare phonon-phonon intrinsic scattering.
Equation~\eqref{eq:energyBTE} at steady state in the ballistic regime can be written as follows
\begin{align}
\bm{v} \cdot \nabla e= 0,
\label{eq:Ballistic}
\end{align}
which indicates that the phonon distribution function inside the system is independent of the physical position $x$.
As shown in~\cref{ballisticpdf,ballisticasymmetric}, for an arbitrary point $P$ inside the system ($| OP| =r$), it can be found that when all phonons with arbitrary directions transporting through $P$, only a portion of them come from the inner hot heat bath with $e^{eq}_R(T_h)$ (between the angle formed by line $P P_1$ and $P P_2$), while the other come from the outer cold heat bath with $e^{eq}_R(T_c)$.
Based on Eqs.~\eqref{eq:T} and~\eqref{eq:energy}, taking an integral of the distribution function in position $P$ over the whole solid angle spaces leads to the temperature at $P(r)$~\cite{li2019,olfe1968}, i.e., (\cref{ballisticasymmetric})
\begin{align}
T(r) = \frac{\beta }{ 2 \pi} T_h + \left(  1- \frac{\beta }{ 2 \pi}   \right) T_c,
\label{eq:TP}
\end{align}
where $\beta=2\arcsin (l/r)$.
Note that if $l/L \rightarrow 1$, $l/r \rightarrow 1$ and $\beta \rightarrow \pi$.
So that Eq.~\eqref{eq:TP} goes to a constant temperature, which is consistent with the results in the ballistic limit in symmetric system~\cite{MajumdarA93Film}.
A comparison of the ballistic phonon transport in symmetric and asymmetric system is shown in~\cref{ballisticasymmetric}.
It can be found that the non-zero temperature gradient in the asymmetric system is built by asymmetric phonon advection, rather than phonon-phonon scattering.

Similarly, the heat flux along the radial direction is
\begin{align}
\bm{q}(r) \cdot \mathbf{n} &=   \int_{0}^{\beta} |\bm{v}| \cos \zeta \left( e^{eq}_R (T_h) -e^{eq}_R (T_c)  \right) d\zeta \notag  \\
&= \frac{2l}{r} |\bm{v}| \left( e^{eq}_R (T_h) -e^{eq}_R (T_c)  \right)   \notag  \\
&= \frac{l}{ \pi r} |\bm{v}| C \left( T_h -T_c  \right) ,
\label{eq:Qballistic}
\end{align}
where $\mathbf{n}$ is the unit normal vector along the radial direction pointing from the inner to the outside.
Then the thermal conductivity along the radial direction in the ballistic limit can be obtained by
\begin{align}
\kappa(r) =- \frac{ \bm{q}(r) \cdot \mathbf{n} }{dT/dr}  =  C |\bm{v}|  \sqrt{\left( r^2-l^2  \right)}.
\label{eq:keffballistic}
\end{align}

As the system size is much larger than the phonon mean free path, i.e., $\text{Kn} \rightarrow 0$, here the phonon transport in the diffusive and phonon hydrodynamic limits~\cite{PhysRevB.99.085202,PhysRev_GK,PhysRev.148.766,Nanalytical,lee_hydrodynamic_2015,cepellotti_phonon_2015} are considered:
\begin{enumerate}
    \item Heat conduction is in the diffusive limit (No N scattering, frequent R scattering).
    \item Heat conduction is in the phonon hydrodynamic limit (No R scattering, frequent N scattering).
\end{enumerate}

When there is no N scattering $\tau_{N}^{-1}=0$ and frequent R scattering happens, the phonon BTE (Eq.\eqref{eq:energyBTE}) becomes
\begin{equation}
\bm{v} \cdot \nabla e= \frac{e^{eq}_{R}-e}{\tau_{R}},
\label{eq:onlyR}
\end{equation}
and the heat conduction is in the diffusive limit.
The distribution function can be approximated
by the first-order Chapman-Enskog expansion as
\begin{equation}
e \approx e^{eq}_{R}  - \tau_{R} \bm{v} \cdot \nabla  e^{eq}_{R}.
\label{eq:fonlyR}
\end{equation}
Then we have
\begin{align}
\bm{q} &=   \int_{2\pi} \bm{v} e  d\Omega  \notag  \\
 & = - \int_{2\pi} \bm{v} \tau_{R} \bm{v} \cdot \nabla  e^{eq}_{R}    d\Omega  \notag\\
 & = - \frac{ \tau_{R} C  |\bm{v}|^2  }{2\pi}  \int_{2\pi}  \bm{s} \bm{s} \cdot \nabla T   d\Omega  \notag\\
 & = - \frac{1}{2} \tau_{R} C |\bm{v}|^2  \nabla T  d\Omega , \notag\\
 & = - \kappa_{\text{diffusive} }   \nabla T ,
\label{eq:qR}
\end{align}
where $\kappa_{\text{diffusive}} = \frac{1}{2} \tau_{R} C |\bm{v}|^2  $.
Due to the energy conservation, the total heat flux $Q(r)$ across the circle with radius $r$ at steady state is a constant, where $Q=2 \pi r \bm{q} \cdot \mathbf{n}$.
Then we have
\begin{align}
\text{constant} &= - 2 \pi r   \kappa_{\text{diffusive}}    dT/dr  \notag   \\
\Longrightarrow \text{constant}  & = - \frac{1}{  dT/d(\ln r)  } \notag   \\
\Longrightarrow  dT &\propto  d(\ln{r} ).
\label{eq:dTdr}
\end{align}

When there is no R scattering $\tau_{R}^{-1}=0$ and frequent N scattering happens, the phonon BTE (Eq.\eqref{eq:energyBTE}) becomes
\begin{equation}
\bm{v} \cdot \nabla e= \frac{e^{eq}_{N}-e}{\tau_{N}},
\label{eq:onlyN}
\end{equation}
and the heat conduction is in the phonon hydrodynamic limit~\cite{PhysRev_GK,PhysRev.148.766,Nanalytical,lee_hydrodynamic_2015,cepellotti_phonon_2015}.
Due to energy and momentum conservation of N scattering~\cite{PhysRev_GK,PhysRev.148.766,Nanalytical}, taking zero- and first- orders of moments of Eq.~\eqref{eq:onlyN} lead to
\begin{align}
\int_{2\pi} \bm{v} \cdot \nabla e  d\Omega &= 0, \label{eq:nzero}  \\
\int_{2\pi} \bm{v} \bm{v} \cdot \nabla e  d\Omega &= 0. \label{eq:nfirst}
\end{align}
A zero-order approximation of the distribution function is made, i.e.,
\begin{equation}
e \approx e^{eq}_{N}.
\label{eq:fonlyN}
\end{equation}
So that Eq.~\eqref{eq:nfirst} becomes
\begin{align}
\int_{2\pi} \bm{v} \bm{v} \cdot \nabla e^{eq}_{N}  d\Omega &= 0,  \\
\int_{2\pi} |\bm{v}|^2  \bm{s} \bm{s} \cdot \nabla \left( \frac{ C(T-T_0)}{2\pi}  +\frac{CT}{2\pi} \frac{ \bm{s} \cdot \bm{u} }{|\bm{v}|}  \right)  d\Omega &= 0, \\
\int_{2\pi} |\bm{v}|^2  \bm{s} \bm{s} \cdot \nabla \left( \frac{ C(T-T_0)}{2\pi}  \right)  d\Omega &= 0, \\
\Longrightarrow \nabla T &=0. \label{eq:nablaTzero}
\end{align}
In other words, in the phonon hydrodynamic limit~\cite{PhysRev_GK,PhysRev.148.766,Nanalytical,lee_hydrodynamic_2015,cepellotti_phonon_2015}, the spatial divergence of temperature is zero.
Namely, in homogeneous 2D disk, the temperature along the radial direction is a constant~\cite{Nanalytical,lee_hydrodynamic_2015,cepellotti_phonon_2015}.

Considering the symmetry, the direction of the drift velocity $\bm{u}$ is along the radial direction, i.e., $|\bm{u}| = \bm{u} \cdot \mathbf{n}$.
When $l<r<L$, the total heat flux $Q(r)$ across the circle with radius $r$ is
\begin{align}
Q(r) &= 2 \pi r  \mathbf{n} \cdot  \bm{q}  \notag   \\
 &= 2 \pi r  \mathbf{n} \cdot  \left( \int_{2\pi}  \bm{v} e^{eq}_N  d\Omega   \right) \notag   \\
 &= 2 \pi r  \mathbf{n} \cdot  \left(  \int_{2\pi} \bm{v} \frac{CT}{2\pi} \frac{ |\bm{K}| \bm{s} \cdot   \bm{u}  }{ \omega } \right) \notag  \\
 & =  CT(r) r  \mathbf{n} \cdot \int_{2\pi} \bm{s} \bm{s} \cdot  \bm{u} d\Omega,\notag  \\
 &= \pi C r  T(r) |\bm{u}(r)| .
\label{eq:QN}
\end{align}
Combining the boundary conditions, i.e., Eq.~\eqref{eq:BC}, when $r \rightarrow l$,
\begin{align}
Q(r) &=   2 \pi l  \mathbf{n} \cdot  \left( \int_{ \bm{s} \cdot \mathbf{n} >0 }  \bm{v} e^{eq}_R (T_h)  d\Omega +  \int_{ \bm{s} \cdot \mathbf{n} <0 }  \bm{v} e^{eq}_N   d\Omega   \right) \notag   \\
 &= 2lC|\bm{v}|T_h-2lC T(r) |\bm{v}|  + \frac{\pi}{2 } C T(r) l  |\bm{u}(r)| .
\label{eq:QNl}
\end{align}
When $r \rightarrow L$,
\begin{align}
Q(r) &=  2 \pi r  \mathbf{n} \cdot  \left( \int_{\bm{s} \cdot \mathbf{n} < 0 }  \bm{v} e^{eq}_R (T_c)   d\Omega  + \int_{\bm{s} \cdot \mathbf{n} >0 }  \bm{v} e^{eq}_N   d\Omega   \right) \notag  \\
 &= -2LC|\bm{v}|T_c  + 2LC |\bm{v}| T(r) + \frac{\pi}{2} CT(r) L  |\bm{u}(r)| .
\label{eq:QNL}
\end{align}
In addition, based on the energy conservation, $Q(r)$ is a constant at steady state.
Combining Eqs.~(\ref{eq:QN},\ref{eq:QNl},\ref{eq:QNL},\ref{eq:nablaTzero}), we have
\begin{align}
T =\frac{lT_h+L T_c } {l+L},
\label{eq:Tcr}
\end{align}
\begin{align}
|\bm{u}(r)  |=  \frac{ 4 |\bm{v}|lL \left( T_h- T_c  \right)    }{  r \pi \left( lT_h+LT_c \right)  }.
\label{eq:uur}
\end{align}

\subsection{3D sphere}

For thermal conduction in 3D concentric ball, the mathematical derivations are similar and straightforward.
Similarly, the analytical solutions in 3D spherical systems can be derived:

In the ballistic limit,
\begin{align}
T &=\frac{T_h}{2}\left( 1-\sqrt{1-l^2/r^2 } \right) +\frac{T_c}{2}\left( 1+\sqrt{1-l^2/r^2 } \right), \label{eq:3DballisticT} \\
\kappa(r) &=\frac{  C |\bm{v}| }{2 }\sqrt{r^2-l^2}. \label{eq:3Dballistick}
\end{align}
In the diffusive limit,
\begin{align}
dT &\propto d(r^{-1}), \label{eq:3DdiffusiveT} \\
\kappa(r) &= \frac{1}{3} C|\bm{v}|^2 \tau_{R} . \label{eq:3Ddiffusivek}
\end{align}
In the phonon hydrodynamic~\cite{PhysRev_GK,PhysRev.148.766,Nanalytical,lee_hydrodynamic_2015,cepellotti_phonon_2015} limit,
\begin{align}
T=\frac{ l^2 T_h +L^2 T_c }{l^2+L^2} . \label{eq:3DhydrodynamicT}
\end{align}

\subsection{Graphene disk}

If considering the phonon dispersion of graphene disk~\cite{pop_varshney_roy_2012,lee_hydrodynamic_2015,cepellotti_phonon_2015}, the analytical solutions are similar to those in 2D disk.
In the ballistic limit,
\begin{align}
T(r)  &= \frac{\beta }{ 2 \pi} T_h + \left(  1- \frac{\beta }{ 2 \pi}   \right) T_c \label{eq:Tdis} \\
\kappa(r) &= \sqrt{\left( r^2-l^2  \right)} \sum_{p} \int \left(|\bm{v}| C \right) d\omega  . \label{eq:kdis}
\end{align}
In the diffusive limit,
\begin{align}
dT &\propto  d(\ln{r} ) ,  \\
\kappa(r) &= \frac{1}{2} \sum_{p} \int  C|\bm{v}|^2 \tau_{R}  d\omega.
\end{align}
In the phonon hydrodynamic limit~\cite{PhysRev_GK,PhysRev.148.766,Nanalytical,lee_hydrodynamic_2015,cepellotti_phonon_2015},
\begin{align}
T =\frac{lT_h+L T_c } {l+L}.
\end{align}

\section{Graded thermal conductivity}

{\color{black}{In this section, the experimental formulas of graded thermal conductivity in 2D disk/3D sphere are used to fit the numerical data approximately.
\begin{table}
\caption{Fitting parameters of Eqs.~\eqref{eq:2dpowerlaws} and~\eqref{eq:Tgraded2D} in 2D disk with fixed system size $L=5l=0.5$ (FIG. 1 in the manuscript). }
\centering
\begin{tabular}{p{2cm}<{\centering} p{2cm}<{\centering} p{2cm}<{\centering} p{4cm}<{\centering} p{4cm}<{\centering} }
\hline
$(\tau_R^{-1},\tau_N^{-1})$ & $\kappa_0$ & $\alpha$  &  $(T_{in} -T_c)/(T_h -T_c)$   &  $(T_{out} -T_c)/(T_h -T_c)$ \\
\hline
$(0.1,10)$  & 4.8 &  5.0  & 0.3 &  0.13      \\
\hline
\end{tabular}
\label{FP2ddisk}
\end{table}
\begin{table}
\caption{Fitting parameters of Eqs.~\eqref{eq:3dpowerlaws} and~\eqref{eq:Tgraded3D} in 3D sphere with fixed system size $L=5l=0.5$ (FIG. 2 in the manuscript). }
\centering
\begin{tabular}{p{2cm}<{\centering} p{2cm}<{\centering} p{2cm}<{\centering} p{4cm}<{\centering} p{4cm}<{\centering} }
\hline
$(\tau_R^{-1},\tau_N^{-1})$ & $\kappa_0$ & $\gamma $  &  $(T_{in} -T_c)/(T_h -T_c)$   &  $(T_{out} -T_c)/(T_h -T_c)$ \\
\hline
$(0.1,10)$  & 0.008  & 5.5   & 0.23 &  0.029   \\
\hline
\end{tabular}
\label{FP3dball}
\end{table}
\begin{table}
\caption{Fitting parameters of Eqs.~\eqref{eq:2dpowerlaws} and~\eqref{eq:Tgraded2D} in graphene disk with fixed system size $L=40~\mu$m, $l=8~\mu$m (FIG. 3(a)(b) in the manuscript). }
\centering
\begin{tabular}{p{2cm}<{\centering} p{2cm}<{\centering} p{2cm}<{\centering} p{4cm}<{\centering} p{4cm}<{\centering} }
\hline
$T_0$ (K) & $\kappa_0$ (W/(m$\cdot$K)) & $\alpha $  &  $(T_{in} -T_c)/(T_h -T_c)$   &  $(T_{out} -T_c)/(T_h -T_c)$ \\
\hline
20  &  11000  & 3.8  & 0.38 &  0.12  \\
\hline
\end{tabular}
\label{FPGDfixedsize}
\end{table}
\begin{table}
\caption{Fitting parameters of Eqs.~\eqref{eq:2dpowerlaws} and~\eqref{eq:Tgraded3D} in graphene disk with fixed temperature $T_0= 300$ K, where $L=5 l$ (FIG. 3(c)(d) in the manuscript). }
\centering
\begin{tabular}{p{2cm}<{\centering} p{2cm}<{\centering} p{2cm}<{\centering} p{4cm}<{\centering} p{4cm}<{\centering} }
\hline
$L$ ($\mu$m) & $\kappa_0$ (W/(m$\cdot$K)) & $\alpha $  &  $(T_{in} -T_c)/(T_h -T_c)$   &  $(T_{out} -T_c)/(T_h -T_c)$ \\
\hline
0.5 &  3000  & 2.0  & 0.53 &  0.16  \\
\hline
\end{tabular}
\label{FPGDfixedT0}
\end{table}

In 2D/graphene disk~\cite{yang_nanoscale_2015,ma_unexpected_2017}, the experimental formula is,
\begin{align}
\kappa (r) &=\kappa_0 \left( R_{2D}^*  \right)^{\alpha}, \quad  &  R_{2D}^* = \frac{ \ln(r/l) -1 }{ \ln(L/l) -1  }  ,  \label{eq:2dpowerlaws}
\end{align}
where $\kappa_0 $ is a constant for a fixed case, $R_{2D}^*$ is the normalized coordination and $\alpha$ is the graded rate~\cite{yang_nanoscale_2015,ma_unexpected_2017}.
In the non-diffusive regime, there is temperature jump near the heat source or heat sink.
To remove this effect, here we ignore the spatial domain near the heat source or sink, for example, $R_{2D}^* \in [0.4, 0.85]$.

Based on the energy conservation, the associated temperature distributions along the radial direction can also be derived based on~\eqref{eq:2dpowerlaws}, i.e.,
\begin{align}
2 \pi r  \kappa \frac{dT}{dr } &= \text{constant},  \notag  \\
\Rightarrow 2 \pi  \kappa \frac{dT}{d \ln(r) } &= \text{constant},  \notag  \\
\Rightarrow \kappa_0 \left( R_{2D}^*  \right)^{\alpha}  \frac{dT}{d \ln(r) } &= \text{constant}, \notag  \\
\Rightarrow \left( R_{2D}^*  \right)^{\alpha} \frac{dT}{d (R_{2D}^*)   } &= \text{constant}.
\label{eq:gradedT1}
\end{align}
When $\alpha \neq 1$, Eq.~\eqref{eq:gradedT1} becomes
\begin{align}
\frac{dT}{d \left( (R_{2D}^*)^{1-\alpha} \right)   } &= \text{constant}.
\label{eq:gradedT2}
\end{align}
When $\alpha = 1$, Eq.~\eqref{eq:gradedT1} becomes
\begin{align}
\frac{dT}{d \left( \ln(R_{2D}^*) \right)   } &= \text{constant}.
\label{eq:gradedT3}
\end{align}

When $R_{2D}^*=0.4$, the temperature is set as $T= T_{in}$, and when $R_{2D}^*=0.85$, the temperature is set as $T= T_{out}$, where $T_{in}$ and $T_{out}$ are two coefficients obtained from phonon BTE.
Then we can get the solutions of temperature along the radial direction based on Eqs.~\eqref{eq:gradedT2} and~\eqref{eq:gradedT3}, i.e.,
\begin{align}
T(R_{2D}^*) &= c_1  (R_{2D}^*)^{1-\alpha}   +c_2 , \quad   \alpha \neq 1,  \label{eq:Tgraded2D}   \\
T(R_{2D}^*) &= c_3  \ln(R_{2D}^*)  +c_4 ,  \quad  \alpha = 1,
\label{eq:2DTgraded}
\end{align}
where
\begin{align}
c_1 &= \frac{T_{out} -T_{in} }{ 0.85^{1-\alpha } - 0.4^{1-\alpha } },  \\
c_2 &= T_{in} -  \frac{T_{out} -T_{in} }{ 2.125^{1-\alpha } - 1  },  \\
c_3 &= \frac{T_{out} -T_{in} }{ \ln 2.125 } ,  \\
c_4 &= T_{in} - \frac{T_{out} -T_{in} }{ \ln 2.125 } * \ln(0.4) .
\end{align}

In 3D sphere, the experimental formula is,
\begin{align}
\kappa (r) &=\kappa_0 \exp \left(  \gamma R_{3D}^*  \right) ,  \quad  & R_{3D}^* = \frac{ 1/l -1/r +1  }{1/l -1/L +1 }  , \label{eq:3dpowerlaws}
\end{align}
where $R_{3D}^*$ is the normalized coordination in 3D sphere and $\gamma$ is a constant for a fixed case.
Here we ignore the spatial domain near the heat source or sink, for example, $R_{3D}^* \in [0.2, 0.8]$.

Based on the energy conservation, the associated temperature distributions along the radial direction can also be derived based on~\eqref{eq:3dpowerlaws}, i.e.,
\begin{align}
4 \pi r^2  \kappa \frac{dT}{dr } &= \text{constant},  \notag  \\
\Rightarrow \kappa_0 \exp \left(  \gamma R_{3D}^*  \right)  \frac{dT}{d (1/r) } &= \text{constant}, \notag  \\
\Rightarrow  \exp \left(  \gamma R_{3D}^*  \right)  \frac{dT}{d (R_{3D}^* ) }  &= \text{constant}, \notag  \\
\Rightarrow   \frac{dT}{d\left(  \exp(-\gamma R_{3D}^*)  \right) }  &= \text{constant}.
\label{eq:gradedT5}
\end{align}
When $R_{3D}^*=0.2$, the temperature is set as $T= T_{in}$, and when $R_{3D}^*=0.8$, the temperature is set as $T= T_{out}$, where $T_{in}$ and $T_{out}$ are two coefficients obtained from phonon BTE.
Then we can obtain the temperature along the radial direction based on Eq.~\eqref{eq:gradedT5}, i.e.,
\begin{align}
T(R_{3D}^*) &= c_5  \exp \left(  - \gamma R_{3D}^*  \right)    +c_6 , \label{eq:Tgraded3D}
\end{align}
where
\begin{align}
c_5 &= \frac{T_{out} -T_{in} }{ \exp \left(  -0.8 \gamma  \right) - \exp \left(  -0.2 \gamma  \right) },  \\
c_6 &= T_{in} -  \frac{T_{out} -T_{in} }{ \exp \left(  -0.6 \gamma  \right) - 1   }.
\end{align}

The detailed fitting parameters of the graded thermal conductivity (FIG. 1-3 in the manuscript) are shown in TABLE~\ref{FP2ddisk},\ref{FP3dball},\ref{FPGDfixedsize},\ref{FPGDfixedT0}, respectively.
}}

\section{Dimensional analysis and numerical results}

We choose $C,~T_0,~L,|\bm{v}|$ as the reference variables to normalize the phonon BTE so that Eq.~\eqref{eq:energyBTE} becomes
\begin{align}
\frac{\bm{v} }{ |\bm{v}|} \cdot \nabla_{\bm{x^*}} e^* = \beta_R (e^{eq,*}_{R}-e^*) + \beta_N (e^{eq,*}_{N}-e^*),
\label{eq:dimensionlessBTEs}
\end{align}
where the dimensionless parameters are
\begin{align}
e^* &=\frac{e}{ C T_0 }, &\quad x^* &=\frac{x}{2L},&\quad e^{eq,*}_R &=\frac{e^{eq}_R }{ C T_0 } , \notag \\
e^{eq,*}_N &=\frac{e^{eq}_N }{ C T_0 },&\quad \beta_N &=\frac{2L}{|\bm{v}| \tau_N},&\quad \beta_R &=\frac{2L}{|\bm{v}| \tau_R},
\label{eq:dimensionlessparameters}
\end{align}
where $2L$ is the diameter of the outer heat bath of 2D disk/3D sphere.
Based on dimensional analysis~\cite{barenblatt1987dimensional}, it can be observed that the heat conduction predicted by Eq.~\eqref{eq:dimensionlessBTEs} is totally decided by two dimensionless parameters, i.e., $\beta_N$ and $\beta_R$, which represent the ratio between the diameter of the outer heat bath to the phonon mean free path of N and R scattering, respectively.
\begin{figure}[htb]
     \centering
     \includegraphics[scale=0.34,clip=true]{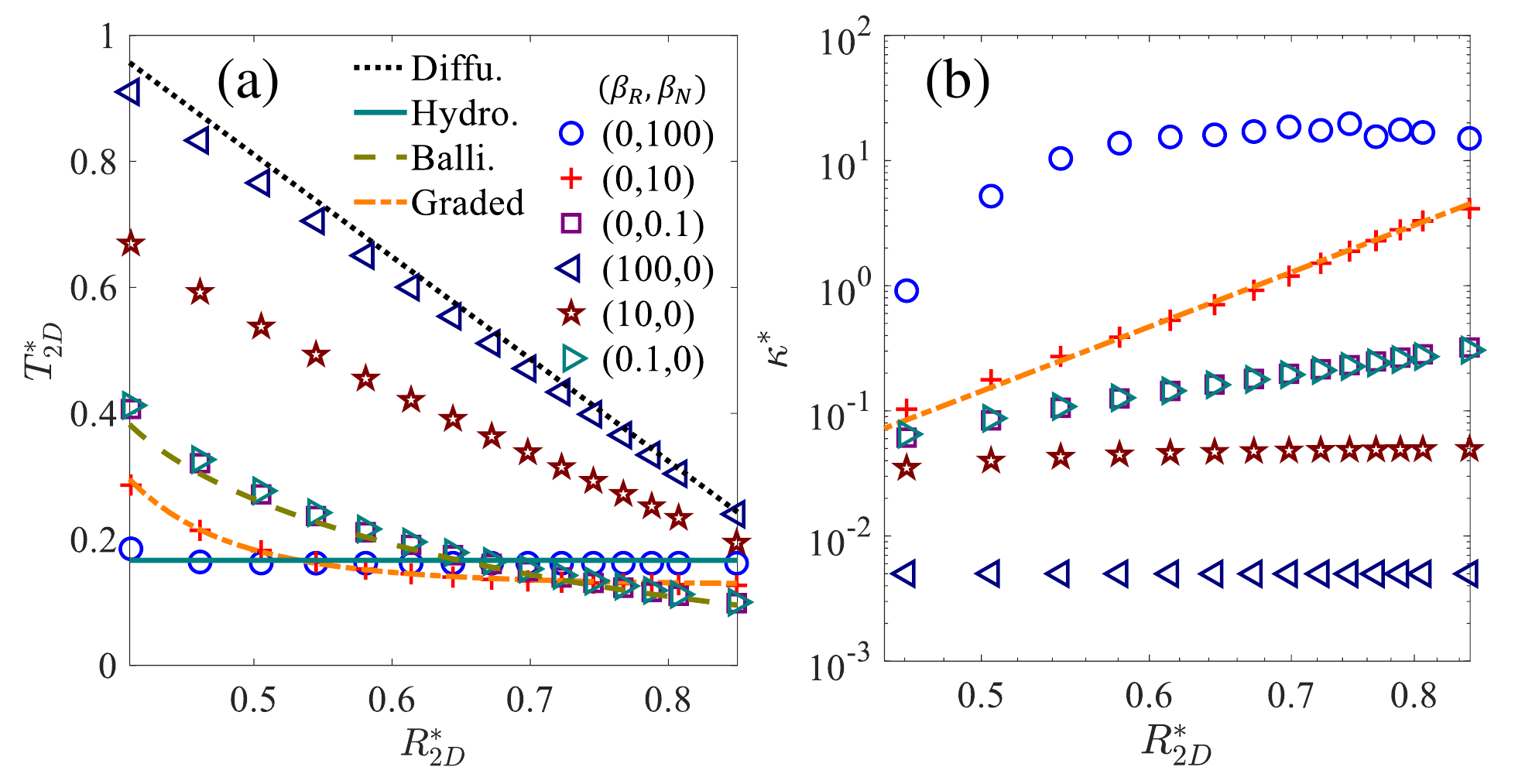}
     \caption{Thermal effects of N and R scattering are investigated individually in 2D disk based on dimensionless BTE~\eqref{eq:dimensionlessBTEs}. (a) Normalized temperature $T_{2D}^*= (T-T_c)/ \Delta T$, $R_{2D}^* = (\ln(r/l)-1) /( \ln(L/l) -1 )$. Only a portion of numerical results are plotted for better observations. Analytical solutions in the ballistic, diffusive and phonon hydrodynamic limits are shown in Eqs.~\eqref{eq:TP},~\eqref{eq:dTdr} and~\eqref{eq:Tcr}, respectively. (b) Normalized thermal conductivity $\kappa^*= \kappa /(2CL|\bm{v}|)$. Symbols are numerical results and the orange dot line are the numerical fittings with power-law functions (Eq.~\ref{eq:2dpowerlawsss}). The detailed fitting parameters can be found in TABLE.~\ref{FP2ddiskmodel}.}
     \label{2DDs}
\end{figure}
\begin{table}
\caption{Fitting parameters of Eqs.~\eqref{eq:2dpowerlawsss} and~\eqref{eq:Tgraded2D} in 2D disk with different $(\beta_R,\beta_N)$ based on Eq.~\eqref{eq:dimensionlessBTEs} (\cref{2DDs}). }
\centering
\begin{tabular}{p{2cm}<{\centering} p{2cm}<{\centering} p{2cm}<{\centering} p{4cm}<{\centering} p{4cm}<{\centering} }
\hline
$(\beta_R ,\beta_N )$ & $\kappa_0$ & $\alpha$  &  $(T_{in} -T_c)/(T_h -T_c)$   &  $(T_{out} -T_c)/(T_h -T_c)$ \\
\hline
(0,10)  & 13.0 &  6.5  & 0.30  &  0.13       \\
\hline
\end{tabular}
\label{FP2ddiskmodel}
\end{table}
\begin{figure}[htb]
    \centering
     \includegraphics[scale=0.34,clip=true]{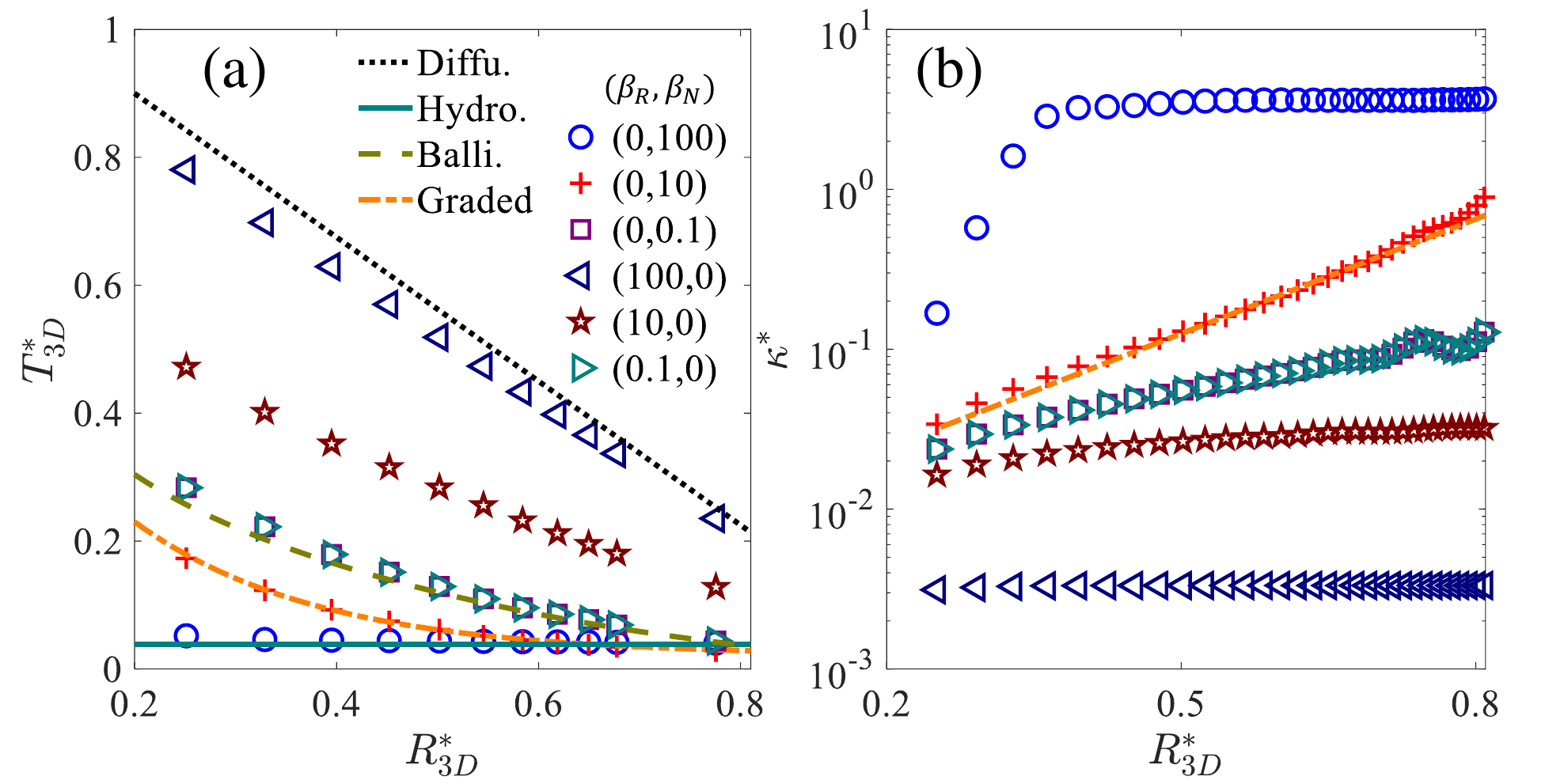}
     \caption{Thermal effects of N and R scattering are investigated individually in 3D sphere based on dimensionless BTE~\eqref{eq:dimensionlessBTEs}. (a) Normalized temperature $T_{3D}^*= (T-T_c)/ \Delta T$, $R_{3D}^* = (1/l -1/r +1 ) / (1/l -1/L +1 ) $. Only a portion of numerical results are plotted for better observations. Analytical solutions in the ballistic, diffusive and phonon hydrodynamic limits are shown in Eqs.~\eqref{eq:3DballisticT},~\eqref{eq:3DdiffusiveT} and~\eqref{eq:3DhydrodynamicT}, respectively. (b) Normalized thermal conductivity $\kappa^*= \kappa /(2CL|\bm{v}|)$. Symbols are numerical results and the orange dot line are the numerical fittings with exponential functions (Eq.~\ref{eq:3dpowerlawsss}).  The detailed fitting parameters can be found in TABLE.~\ref{FP3dballmodel}.}
     \label{3DDs}
\end{figure}
\begin{table}
\caption{Fitting parameters of Eqs.~\eqref{eq:3dpowerlawsss} and~\eqref{eq:Tgraded3D} in 3D sphere with different $(\beta_R,\beta_N)$ based on Eq.~\eqref{eq:dimensionlessBTEs} (\cref{3DDs}). }
\centering
\begin{tabular}{p{2cm}<{\centering} p{2cm}<{\centering} p{2cm}<{\centering} p{4cm}<{\centering} p{4cm}<{\centering} }
\hline
$(\beta_R ,\beta_N )$ & $\kappa_0$ & $\gamma $  &  $(T_{in} -T_c)/(T_h -T_c)$   &  $(T_{out} -T_c)/(T_h -T_c)$ \\
\hline
(0,10)  & 0.008 &  5.5  & 0.23  &  0.029      \\
\hline
\end{tabular}
\label{FP3dballmodel}
\end{table}

The thermal effects of N scattering ($\beta_N$) and R scattering ($\beta_R$) are investigated individually based on Eq.~\eqref{eq:dimensionlessBTEs}.
In simulation of 2D disk~(\cref{2DDs}) and 3D sphere~(\cref{3DDs}), the Debye approximation and gray model~\cite{ChenG05Oxford} are used, where no phonon dispersion and polarization are considered.
Numerical results are consistent with the analytical solutions in the ballistic, diffusive and phonon hydrodynamic~\cite{PhysRev_GK,PhysRev.148.766,Nanalytical,lee_hydrodynamic_2015,cepellotti_phonon_2015} limits.
Furthermore, it can be observed that as $\beta_N$ and $\beta_R$ are small, the thermal conductivity along the radial direction is not a constant.
Similarly, the numerical results (\cref{2DDs,3DDs}) of graded thermal conductivity in 2D disk/3D sphere are fitted with experimental formulas, i.e.,
\begin{align}
\kappa^* &=\kappa_0 \left( R_{2D}^*  \right)^{\alpha},   \label{eq:2dpowerlawsss}  \\
\kappa^* &=\kappa_0 \exp \left(  \gamma R_{3D}^*  \right) ,  \label{eq:3dpowerlawsss}
\end{align}
where the normalized thermal conductivity is $\kappa^* = \kappa /(2CL|\bm{v}|)$.
The detailed fitting parameters of the graded thermal conductivity in 2D disk/3D sphere are shown in TABLE~\ref{FP2ddiskmodel} and~\ref{FP3dballmodel}, respectively.

\bibliography{phonon}